\documentclass[aip,jcp,amsmath,amssymb,graphicx,reprint]{revtex4-2}

\usepackage{graphicx}
\usepackage{dcolumn}
\usepackage{bm}
\usepackage{physics}
\usepackage[utf8]{inputenc}
\usepackage[T1]{fontenc}
\usepackage{mathptmx}
\usepackage{xcolor}
\usepackage{array}
\usepackage[title]{appendix}
\newcolumntype{L}[1]{>{\raggedright\let\newline\\\arraybackslash\hspace{0pt}}m{#1}}
\newcolumntype{C}[1]{>{\centering\let\newline\\\arraybackslash\hspace{0pt}}m{#1}}
\newcolumntype{R}[1]{>{\raggedleft\let\newline\\\arraybackslash\hspace{0pt}}m{#1}}

\begin{document}

\title{A novel non-adiabatic spin relaxation mechanism in molecular qubits}

\author{Philip Shushkov}
\email[]{phgshush@iu.edu}
\affiliation{Department of Chemistry, Indiana University, Bloomington, Indiana, 47405}

\date{\today}

\begin{abstract}
The interaction of electronic spin and molecular vibrations mediated by spin-orbit coupling governs spin relaxation in molecular qubits. I derive an extended molecular spin Hamiltonian that includes both adiabatic and non-adiabatic spin-dependent interactions, and I implement the computation of its matrix elements using state-of-the-art density functional theory. The new molecular spin Hamiltonian contains a novel spin-vibrational orbit interaction with non-adiabatic origin together with the traditional molecular Zeeman and zero-field splitting interactions with adiabatic origin. The spin-vibrational orbit interaction represents a non-Abelian Berry curvature on the ground-state electronic manifold and corresponds to an effective magnetic field in the electronic spin dynamics. I further develop a spin relaxation rate model that estimates the spin relaxation time via the two-phonon Raman process. An application of the extended molecular spin Hamiltonian together with the spin relaxation rate model to Cu(II) porphyrin, a prototypical $S=1/2$ molecular qubit, demonstrates that the spin relaxation time at elevated temperatures is dominated by the non-adiabatic spin-vibrational orbit interaction. The computed spin relaxation rate and its magnetic field orientation dependence are in excellent agreement with experimental measurements.
\end{abstract}


\maketitle

\section{Introduction}
Molecular qubits, paramagnetic systems that exhibit long spin-coherence times, are emerging as a promising platform for the implementation of quantum information processing.\cite{gaita2019molecular,fataftah2018progress,atzori2019second,graham2017forging} The molecular electronic spin is an excellent candidate to encode quantum information because of protection by time-reversal symmetry. Taken together with the unprecedented accuracy of chemical methods to synthesize and assemble molecular systems, there has been a recent surge in efforts to engineer molecular qubits that are suitable to advance the much sought-after room-temperature quantum technologies.\cite{kazmierczak2023t,kazmierczak2022illuminating,kazmierczak2021impact,zadrozny2015millisecond,amdur2022chemical,amdur2022chemical,yu2020spin,fataftah2019metal,fataftah2020trigonal,graham2014influence,follmer2020understanding,tesi2016quantum,atzori2016room,atzori2018structural,atzori2017spin}

A major limiting factor to the room-temperature quantum information storage in molecular qubits is the spin relaxation time $T_1$, called also spin-lattice or longitudinal relaxation time,\cite{goldman2001formal} that characterizes the timescale of thermal equilibration of the electronic spin by the molecular vibrational motion.\cite{amdur2022chemical,kazmierczak2022illuminating,kazmierczak2021impact,kazmierczak2023t} The spin relaxation theory dates back to the pioneering work of Van Vleck,\cite{van1940paramagnetic} Mattuck and Strandberg,\cite{mattuck1960spin} and Orbach,\cite{orbach1961spin} who focused on the relaxation dynamics of paramagnetic impurities in crystals. The physical picture that emerged from their seminal contributions is that both adiabatic and non-adiabatic processes can dominate the spin relaxation time depending on the specifics of the electronic structure of the paramagnetic impurities. Furthermore, they showed that one-phonon processes contribute to the the spin relaxation dynamics only at very low temperatures and at elevated temperatures spin relaxation is driven by two-phonon absorption-emission processes.

Density functional theory and multi-reference wavefunction approaches have provided a firm theoretical basis to predict molecular spin interactions.\cite{neese2001prediction,neese1998calculation,neese2006importance,neese2005efficient,singh2018challenges,gauss2009calculation} Fundamental to this success is the concept of the static spin Hamiltonian, an effective Hamiltonian defined at the equilibrium nuclear geometry that incorporates the influence of the molecular electronic structure in spin-dependent interactions.\cite{mcweeny1965origin,pryce1950modified} The dynamical extension of the static spin Hamiltonian by accounting for the nuclear geometry-dependence of the traditional spin-dependent interactions has underscored recent efforts to simulate the spin relaxation dynamics of molecular qubits.\cite{lunghi2020limit,lunghi2020multiple,lunghi2022toward,briganti2021complete,garlatti2023critical,albino2019first,reta2021ab,kragskow2023spin} This approach to spin relaxation, however, does not account for the contribution of the non-adiabatic interactions.

The goal of the paper is to derive a molecular spin Hamiltonian that includes both adiabatic and non-adiabatic spin-dependent interactions and to implement the computation of the matrix elements of this Hamiltonian using density functional theory. I achieve this goal by applying the Born-Oppenheimer approximation in the adiabatic representation for the electronic wavefunctions,\cite{yarkony1996diabolical,mead1992geometric} circumventing the need for diabatization, and allowing seamless integration with linear response density functional theory. Similar to the static spin Hamiltonian, I use unitary degenerate perturbation theory\cite{van1929sigma,primas1963generalized} to derive the molecular spin Hamiltonian, which limits its applicability to molecular systems with weak spin-orbit interaction and orbitally non-degenerate ground electronic states.\cite{pryce1950modified} The derived molecular spin Hamiltonian contains a novel, non-adiabatic spin-vibrational orbit interaction together with the traditional molecular Zeeman and zero-field splitting interactions\cite{mcweeny1965origin} that have an adiabatic origin. I further develop a rate model to estimate the contribution to the spin relaxation time of the interactions in the molecular spin Hamiltonian. The rate model is specialized to elevated temperatures and evaluates the spin relaxation time via the two-phonon Raman process\cite{van1940paramagnetic} using density functional calculations on the isolated paramagnetic molecule. A pilot application of the molecular spin Hamiltonian together with the spin relaxation rate model to Cu(II) porphyrin,\cite{du1996electron} a prototypical $S=1/2$ molecular qubit,\cite{yu2019concentrated,moisanu2023crystalline} demonstrates that the two-phonon spin relaxation time is dominated by the non-adiabatic spin-vibrational orbit interaction with doubly degenerate normal modes being the major vibrational relaxation channel.

The paper is organized as follows: I present the derivation of the molecular spin Hamiltonian and of the spin relaxation rate model in Sec.~\ref{theory}. I outline the numerical evaluation of the molecular spin Hamiltonian matrix elements using density functional theory in Sec.~\ref{dft}. I present the results of the application of the new approach to Cu(II) porphyrin in Sec.~\ref{results}, and I conclude the paper with an outline of future directions in Sec.~\ref{conclusions}.

\section{Theory}\label{theory}
Spin-vibrational interactions involve the coupling of the electronic spin with the molecular vibrations that is mediated by the electronic orbital motion. I start Sec.~\ref{theory} by specifying the molecular Hamiltonian and establishing the notation used in the paper. I then present an overview of unitary degenerate perturbation theory\cite{primas1963generalized} that I apply to derive the molecular spin Hamiltonian in Sec.~\ref{molspinHam} and proceed with the application of the Born-Oppenheimer approximation. In Sec.~\ref{molspinHam}, I derive the expression for the Berry connection\cite{berry1984quantal,mead1979determination,shapere1989geometric} on the ground-state spin manifold and arrive at a molecular spin Hamiltonian that contains a novel spin-vibronic vector potential. In Sec.~\ref{gauge}, I apply a unitary gauge transformation\cite{mead1992geometric} to a symmetrized gauge for the spin-vibronic vector potential and derive the spin-vibrational orbit interaction. Harmonic expansion of the resulting molecular spin Hamiltonian in Sec.~\ref{harmonic} gives one- and two-phonon spin-vibrational interactions with non-adiabatic and adiabatic origins. In Sec.~\ref{spinrate}, I present a rate model for the computation of the spin relaxation time of $S=1/2$ molecular qubits at elevated temperature and at weak to moderate magnetic field intensities using the two-phonon spin-vibrational interactions of Sec.~\ref{harmonic}.

\subsection{Molecular Spin Hamiltonian}\label{molspinHam}
I write the molecular Hamiltonian,\cite{mcweeny1965origin} including relativistic interactions and the interaction of the molecule with an external magnetic field as:
\begin{equation}\label{totham}
    \hat{H}_{\text{molec}} = \hat{T}_{\textbf{R}} + \hat{H}^{\text{NR}}_{\textbf{R}} + \hat{H}^{\text{SOZ}}_{\textbf{R}}.
\end{equation}
$\hat{T}_{\textbf{R}}$ in Eq.~(\ref{totham}) is the nuclear kinetic energy, $\hat{H}^{\text{NR}}_{\textbf{R}}$ - the non-relativistic electronic Hamiltonian together with the spin-independent, scalar relativistic corrections, and $\hat{H}^{\text{SOZ}}_{\textbf{R}}$ - the spin-dependent relativistic interactions that include the spin-orbit, the electronic Zeeman, and the electronic spin-spin interactions. $\textbf{R}$ stands for the collection of Cartesian nuclear coordinates, and the subscript denotes the explicit dependence of the Hamiltonian terms on the nuclear configuration. In this paper, I do not include the interactions of the electronic spin with the nuclear magnetic moments, which give rise to hyperfine splitting of the electronic energy levels.

I define the zero-order adiabatic electronic wavefunctions, $\ket{K^{(0)}_{\textbf{R}}}$, and the zero-order adiabatic electronic potential energy surfaces, $E^{(0)}_{K,\textbf{R}}$, as the eigenvectors and the eigenvalues of the non-relativistic electronic Hamiltonian:
\begin{equation}\label{eigenv}
    \hat{H}^{\text{NR}}_{\textbf{R}} \ket{K^{(0)}_{\textbf{R}}} = E^{(0)}_{K,\textbf{R}} \ket{K^{(0)}_{\textbf{R}}}.
\end{equation}
The eigenfunctions and eigenvalues are zero order with respect to the spin-dependent interactions, in which case the zero-order electronic states form multiplets of degenerate states that correspond to different projection of the electronic spin angular momentum. To account for the degeneracy of the zero-order electronic spectrum, the multi-index $K$ stands for a collection of quantum numbers $K=\{kSM_S\}$ that specify the orbital quantum number $k$, the spin quantum number $S$, and the associated spin projection $M_S$. I assume that the state degeneracy in the multiplets is only of spin origin and exclude degeneracy in the multiplets that results from spatial symmetry. The adiabatic eigenfunctions in Eq.~(\ref{eigenv}) are parametrically dependent on the nuclear configuration, $\textbf{R}$, and form an orthonormal set at each nuclear geometry. 

The spin-dependent interactions in Eq.~(\ref{totham}) couple the electronic states of the non-relativistic Hamiltonian and remove partially or fully the degeneracy of the electronic spectrum. I apply unitary degenerate perturbation theory\cite{primas1963generalized} to diagonalize the inter-multiplet interactions due to the spin-dependent Hamiltonian $\hat{H}^{\text{SOZ}}_{\textbf{R}}$. The perturbative treatment is justified for molecules that consist of atoms of light elements, including early transition metal and main group elements, because for them the spin-dependent multiplet splittings are significantly smaller than the energy differences between the zero-order electronic multiplets. The perturbed wavefunctions, $\ket{K_{\textbf{R}}}$, in unitary perturbation theory are obtained by a unitary transformation of the zero-order wavefunctions: 
\begin{equation}\label{perturbedwfn}
    \ket{K_{\textbf{R}}} = e^{\hat{G}_{\textbf{R}}} \ket{K^{(0)}_{\textbf{R}}},
\end{equation}
where $\hat{G}$ is the generator of the transformation and satisfies $\hat{G}^{\dagger} = -\hat{G}$ to ensure the unitarity of the transformation. The generator is obtained perturbatively from the requirement of vanishing inter-multiplet interactions in the unitarily transformed Hamiltonian, and the perturbed eigenvalues derive from the intra-multiplet blocks of the transformed Hamiltonian:
\begin{equation}\label{transformH}
   \bra{J^{(0)}_{\textbf{R}}} e^{-\hat{G}_{\textbf{R}}} \left( \hat{H}^{\text{NR}}_{\textbf{R}} + \hat{H}^{\text{SOZ}}_{\textbf{R}} \right) e^{\hat{G}_{\textbf{R}}} \ket{K^{(0)}_{\textbf{R}}} = \delta_{JK} E_{K'K,\textbf{R}}.
\end{equation}
Multi-indeces $J,K,..$ refer to general multiplets, and I reserve the multi-index $I$ for the ground-state multiplet. $H_{IJ}$ stands for the matrix block between two different multiplets $I$ and $J$, whereas $H_{I'I}$ stands for the matrix block of the multiplet $I$. $\delta_{JK}$ stands for a multi-index Kronecker delta symbol. Use of the Baker-Campbell-Hausdorff formula in Eq.~(\ref{transformH}) and collection of terms according to the order of perturbation provides the equation for the first-order generator:
\begin{equation}\label{firstorderG}
    \left[ \hat{H}^{\text{NR}}_{\textbf{R}} , \hat{G}^{(1)}_{\textbf{R}} \right]_{JK} + H^{\text{SOZ}}_{JK,\textbf{R}} = 0,
\end{equation}
which is solved for the inter-multiplet blocks of the generator $G^{(1)}_{JK}$ and is supplemented with the additional condition of vanishing intra-multiplet blocks $G^{(1)}_{K'K}=0$. The perturbation expansion in Eq.~(\ref{transformH}) gives also the perturbed eigenvalues up to second order in the perturbation:
\begin{equation}\label{secondorderE}
    E_{K'K,\textbf{R}} = E^{(0)}_{K,\textbf{R}} \delta_{K'K} + H^{\text{SOZ}}_{K'K,\textbf{R}} + \frac{1}{2} \left[ \hat{H}^{\text{SOZ}}_{\textbf{R}} , \hat{G}^{(1)}_{\textbf{R}} \right]_{K'K}.
\end{equation}
Eq.~(\ref{firstorderG}) gives the ground-state electronic wavefunction to first order in the spin-orbit perturbation as:
\begin{equation}\label{expansionwfn}
\begin{split}
    \ket{I_{\textbf{R}}} &= \ket{I^{(0)}_{\textbf{R}}} + \hat{G}^{(1)}_{\textbf{R}} \ket{I^{(0)}_{\textbf{R}}} \\ &= \ket{I^{(0)}_{\textbf{R}}} - \sum_{J \ne I} \ket{J^{(0)}_{\textbf{R}}} \frac{H^{\text{SOI}}_{JI,\textbf{R}}}{E^{(0)}_{J,\textbf{R}}-E^{(0)}_{I,\textbf{R}}}.
\end{split}
\end{equation}
Components of multiplets with different projections of spin angular momentum are non-interacting at zero order; coupling between them arises in the first-order contribution to the electronic wavefunction, the second term in Eq.~(\ref{expansionwfn}), from the spin-orbit interaction (SOI) in $\hat{H}^{\text{SOZ}}_{\textbf{R}}$. The perturbative expansion of the ground-state multiplet energies from Eq.~(\ref{secondorderE}) is given by:
\begin{equation}\label{expansion}
\begin{split}
    E_{I'I,\textbf{R}} &= E^{(0)}_{I,\textbf{R}} \delta_{I'I} + E^{(1)}_{I'I,\textbf{R}} + E^{(2)}_{I'I,\textbf{R}} \\ &= E^{(0)}_{I,\textbf{R}} \delta_{I'I} + H^{\text{SOZ}}_{I'I,\textbf{R}} + \sum_{J \ne I} \frac{H^{\text{SOZ}}_{I'J,\textbf{R}} \; H^{\text{SOZ}}_{JI,\textbf{R}}}{E^{(0)}_{I,\textbf{R}}-E^{(0)}_{J,\textbf{R}}},
\end{split}
\end{equation}
where $E_{I'I,\textbf{R}}$ is a matrix within the ground-state manifold, $E_{M'M,\textbf{R}}$. The first-order and second-order adiabatic energy contributions, the second and third term in Eq.~(\ref{expansion}), give rise to the traditional molecular spin Hamiltonian with the molecular Zeeman interaction, $\mu_B \vec{B} \textbf{g}_{\textbf{R}} \vec{S}$, characterized by the electronic g-tensor and $\mu_B$ the Bohr magneton, and the zero-field splitting interaction, $\vec{S} \textbf{D}_{\textbf{R}} \vec{S}$, characterized by the electronic D-tensor. Both the g-tensor and D-tensor depend parametrically on the nuclear positions, and the existence of the molecular spin Hamiltonian requires an orbitally-non-degenerate ground state, which is the case for all systems that I consider in the paper. I establish the connection with the traditional spin Hamiltonian and the spin-component structure of the adiabatic wavefunctions in Eq.~(\ref{expansionwfn}) using the Wigner-Eckart theorem in Appendix~\ref{appendixa}.  

I apply the Born-Oppenheimer approximation\cite{mead1992geometric,yarkony1996diabolical} to \textit{uncouple} the electron-nuclear dynamics on \textit{different} electronic multiplets and preserve the \textit{coupled} electron-nuclear dynamics \textit{within} an electronic multiplet. This treatment of the electron-nuclear dynamics is justified when the energy differences \textit{between} electronic multiplets are significantly larger than the vibrational quanta involved in the spin relaxation dynamics, whereas the energy differences \textit{within} electronic multiplets are of similar magnitude or smaller than the vibrational quanta of the nuclear dynamics. This regime holds for the $S=1/2$ systems considered in this work. The case where two or more electronic multiplets come close in energy and interact non-adiabatically as it occurs in conical intersections is outside the scope of the present treatment.
In the Born-Oppenheimer approximation, the vibronic wavefunction for the ground-state multiplet is expressed as the sum of products of the electronic wavefunction of a multiplet component, $\ket{iSM,\textbf{R}}$, and the associated nuclear wavefunction, $\chi_{iSM}(\textbf{R})$:
\begin{equation}\label{bowfn}
    \ket{\Psi^{\text{BO}}_I} = \sum_{M, M'} \ket{iSM',\textbf{R}} c_{M'M,\textbf{R}_0} \; \chi_{iSM}(\textbf{R}) = \ket{I_{\textbf{R}}} \chi_{I}(\textbf{R}).
\end{equation}
The electronic wavefunctions in Eq.~(\ref{bowfn}) are the perturbed wavefunctions of Eqs.~(\ref{perturbedwfn}) and (\ref{expansionwfn}), and the last equality implies a sum over the multiplet components. I choose a diabatic representation within the multiplet manifold that conserves the character of the multiplet components, the dominant projection of spin angular momentum $M$, and the matrix $c_{M'M,\textbf{R}_0}$ in Eq.~(\ref{bowfn}) allows for a specific choice of the spin quantization axis at a fixed nuclear configuration. I use this matrix to specify the reference spin quantization axis for the spin relaxation dynamics at the equilibrium nuclear configuration. If the matrix $c_{M'M,\textbf{R}}$ were allowed to vary with nuclear position and at each nuclear configuration were chosen as the eigenvectors of the matrix, $E_{I'I,\textbf{R}}$, in Eq.~(\ref{expansion}), then the resulting wavefunctions, $\sum_{M, M'} \ket{iSM',\textbf{R}} c_{M'M,\textbf{R}}$, would be the adiabatic electronic wavefunctions of the multiplet components and the associated eigenvalues would be the adiabatic potential energy surfaces of the multiplet components.  

To derive the Born-Oppenheimer Hamiltonian for the ground-state multiplet, I apply the nuclear kinetic energy operator to the Born-Oppenheimer wavefunction in Eq.~(\ref{bowfn}) and obtain:
\begin{equation}\label{kinen}
\begin{split}
    \hat{T}_{\textbf{R}} \ket{\Psi^{\text{BO}}_I} = &\ket{I_{\textbf{R}}} \hat{T}_{\textbf{R}} \chi_{I}(\textbf{R}) 
    \\ &-i \partial_{\mu} \ket{I_{\textbf{R}}} \hat{p}_{\mu} \chi_{I}(\textbf{R}) +
    \chi_{I}(\textbf{R}) \hat{T}_{\textbf{R}} \ket{I_{\textbf{R}}}.
\end{split}
\end{equation}
The notation in Eq.~(\ref{kinen}) implies a sum over the components of the electronic multiplet. I use the Einstein convention for an unrestricted summation over repeated indeces, and I explicitly include the sum symbol when there are restrictions on the sum indeces and when there is a summation without index repetition. $\mu$ is the index of the mass-weighted nuclear Cartesian coordinates, and $\hat{p}_{\mu}$ is the nuclear momentum operator of coordinate $\mu$. $i$ is the imaginary unit, and I use atomic units, such that $\hbar=1$. The action of the nuclear kinetic energy operator gives terms with nuclear derivatives acting only on the nuclear, first term, and the electronic wavefunctions, third term, as well as a mixed term with nuclear derivatives acting on both the electronic and nuclear wavefunctions.   

I project the molecular Hamiltonian, Eq.~(\ref{totham}), in the ground-state multiplet, Eq.~(\ref{bowfn}), and use Eqs.~(\ref{transformH}) and (\ref{kinen}) to derive the Born-Oppenheimer molecular Hamiltonian: 
\begin{equation}\label{BOHam}
    \hat{H}^{BO}_{I'I} = \delta_{I'I} \; \hat{T}_{\textbf{R}}
    -i A_{I'I\mu,\textbf{R}} \; \hat{p}_{\mu} +
    D_{I'I,\textbf{R}} + E_{I'I,\textbf{R}},
\end{equation}
The Hamiltonian in Eq.~(\ref{BOHam}) is a matrix within the ground-state electronic multiplet space and an operator in nuclear Cartesian coordinate space. The first term is the nuclear kinetic energy, and the last term is the adiabatic potential energy matrix of the ground-state multiplet, Eq.~(\ref{expansion}). The second and third terms are non-adiabatic interactions and contain the first-derivative non-adiabatic coupling matrix,
\begin{equation}\label{Berryconn}
    A_{I'I\mu,\textbf{R}}=\bra{I'_{\textbf{R}}} \partial_{\mu} \ket{I_{\textbf{R}}},
\end{equation}
and the second-derivative non-adiabatic coupling matrix,
\begin{equation}\label{Dpot}
    D_{I'I,\textbf{R}}=\bra{I'_{\textbf{R}}} \hat{T}_{\textbf{R}} \ket{I_{\textbf{R}}}.
\end{equation}
$A_{I'I\mu,\textbf{R}}$ is the Berry connection,\cite{berry1984quantal,shapere1989geometric,mead1992geometric,mead1987molecular} which in this case has a non-Abelian algebraic structure and arises as a result of non-adiabatic interactions with excited electronic multiplets. All inter-multiplet non-adiabatic coupling matrices vanish because of the application of the Born-Oppenheimer approximation. 

I expand the Berry connection in the spin-orbit interaction to derive explicit expressions for the non-adiabatic coupling matrix elements between multiplet components. I use the perturbative expansion of the electronic wavefunctions, Eq.~(\ref{perturbedwfn}), and the Baker-Campbell-Hausdorff formula in Eq.~(\ref{Berryconn}) and obtain to first order in the SOI:
\begin{equation}\label{Vpot1}
\begin{split}
    A_{I'I\mu,\textbf{R}} &= \bra{I'^{(0)}_{\textbf{R}}} e^{- \hat{G}_\textbf{R}} \partial_{\mu} e^{\hat{G}_\textbf{R}} \ket{I^{(0)}_{\textbf{R}}} \\ &= \bra{I'^{(0)}_{\textbf{R}}} \ket{\partial_{\mu} I^{(0)}_{\textbf{R}}} + \bra{I'^{(0)}_{\textbf{R}}} \left[ \partial_{\mu}, \hat{G}^{(1)}_\textbf{R} \right] \ket{I^{(0)}_{\textbf{R}}}.
\end{split}
\end{equation}
The zero-order term of the Berry connection $A^{(0)}_{I'I\mu,\textbf{R}}$, the first term in the last equality in Eq.~(\ref{Vpot1}), is diagonal in the ground-state multiplet space and is the generalization of the traditional Berry connection that arises from conical intersections.\cite{mead1979determination,berry1984quantal} Because I consider systems away from conical intersections in this paper, $A^{(0)}_{I'I\mu,\textbf{R}}$ vanishes. The first-order term in the spin-orbit interaction, $A^{(1)}_{I'I\mu,\textbf{R}}$, the second term in Eq.~(\ref{Vpot1}), is the leading contribution to the Berry connection on the ground-state manifold and it couples different multiplet components as a result of the interplay of spin-orbit and non-adiabatic interactions with excited electronic states. Expansion of the commutator in Eq.~(\ref{Vpot1}) gives an explicit formula for the first-order Berry connection in terms of zero-order adiabatic electronic wavefunctions:
\begin{equation}\label{Vpot3}
\begin{split}
    A^{(1)}_{I'I\mu,\textbf{R}} = \sum_{J \ne I}  \frac{H^{\text{SOI}}_{I'J,\textbf{R}}}{E^{(0)}_{I',\textbf{R}}-E^{(0)}_{J,\textbf{R}}} \bra{J^{(0)}_{\textbf{R}}} \ket{\partial_{\mu} I^{(0)}_{\textbf{R}}} \\
    -\sum_{J \ne I} \bra{I'^{(0)}_{\textbf{R}}} \ket{\partial_{\mu} J^{(0)}_{\textbf{R}}} \frac{H^{\text{SOI}}_{JI,\textbf{R}}}{E^{(0)}_{J,\textbf{R}}-E^{(0)}_{I,\textbf{R}}}.
\end{split}
\end{equation}
The non-Abelian algebraic structure of $A^{(1)}_{I'I\mu,\textbf{R}}$ follows from an application of the Wigner-Eckart theorem\cite{sakurai2014modern} to Eq.~(\ref{Vpot3}), which gives an expression for $A^{(1)}_{I'I\mu,\textbf{R}}$ in terms of the set of non-commuting spin matrices, $S_{M'M,\alpha}$, with dimension equal to the dimensionality of the ground-state multiplet space:
\begin{equation}\label{Vpot4}
\begin{split}
    A^{(1)}_{I'I\mu,\textbf{R}} &= i S_{M'M,\alpha} a^{(1)}_{\alpha \mu,\textbf{R}}, \\ 
    a^{(1)}_{\alpha \mu,\textbf{R}} &= 2 \sum_{j \ne i} \frac{h^{\text{SOI}}_{ij\alpha,\textbf{R}}}{E^{(0)}_{i,\textbf{R}}-E^{(0)}_{j,\textbf{R}}} \bra{j^{(0)}_{\textbf{R}}} \ket{\partial_{\mu} i^{(0)}_{\textbf{R}}}.
\end{split}
\end{equation}
The sum over the Cartesian index $\alpha=x,y,z$ is implicit in Eq.~(\ref{Vpot4}), and $h^{\text{SOI}}_{ij\alpha,\textbf{R}}$ is the imaginary part of the one-electron, mean-field spin-orbit coupling operator as given in Appendix~\ref{appendixa}. The functions $a^{(1)}_{\alpha \mu,\textbf{R}}$ are the three Cartesian components of a vector potential associated with each nuclear degree of freedom $\mu$. $a^{(1)}_{\alpha \mu,\textbf{R}}$ are real functions of the nuclear positions, which ensures that $A^{(1)}_{I'I\mu,\textbf{R}}$ is an anti-Hermitian matrix. Note that with the definition of the Berry connection in Eq.~(\ref{Berryconn}), the product $-i\hbar A^{(1)}_{I'I\mu,\textbf{R}}$ is a Hermitian matrix. The last equality of Eq.~(\ref{Vpot4}) expresses the first-order vector potential, $a^{(1)}_{\alpha \mu,\textbf{R}}$, in terms of matrix elements of zero-order adiabatic electronic wavefunctions, $\ket{\{jSS\}^{(0)}_{\textbf{R}}}=\ket{j^{(0)}_{\textbf{R}}}$, with spin quantum number $S$ equal to the spin quantum number of the ground-state multiplet and maximum projection of the spin angular momentum $M_S=S$. $\ket{i^{(0)}_{\textbf{R}}}$ is the maximum-spin-projection ground-state electronic wavefunction and the sum in Eq.~(\ref{Vpot4}) runs over the maximum-spin-projection excited-state electronic wavefunctions. The benefit of the Wigner-Eckart theorem is to separate the exact spin-component structure from the electronic matrix elements and to express all relevant quantities in terms of electronic wavefunctions of fixed projection of spin angular momentum. I implement Eq.~(\ref{Vpot4}) together with state-of-the-art density functional theory approximations to evaluate the first-order Berry connection in Sec.~\ref{dft}.

Similar expansion of the second-derivative non-adiabatic coupling matrix, Eq.~(\ref{Dpot}), in the spin-orbit interaction gives:
\begin{equation}\label{Dpot1}
\begin{split}
    D_{I'I,\textbf{R}} &= \bra{I'^{(0)}_{\textbf{R}}}  \hat{T}_{\textbf{R}} \ket{I^{(0)}_{\textbf{R}}} + \bra{I'^{(0)}_{\textbf{R}}} \left[ \hat{T}_{\textbf{R}}, \hat{G}^{(1)}_\textbf{R} \right] \ket{I^{(0)}_{\textbf{R}}} \\ &= D^{(0)}_{I'I,\textbf{R}} + D^{(1)}_{I'I,\textbf{R}}.
\end{split}
\end{equation}
The zero-order term of the second-derivative non-adiabatic coupling matrix, $D^{(0)}_{I'I,\textbf{R}}$, is similarly diagonal in the ground-state multiplet space and vanishes away from conical intersections. The first-order contribution, $D^{(1)}_{I'I,\textbf{R}}$, can be entirely expressed in terms of the first-order Berry connection:
\begin{equation}\label{Dpot2}
\begin{split}
    D^{(1)}_{I'I,\textbf{R}} &= - \sum_\mu \frac{1}{2} \bra{I'^{(0)}_{\textbf{R}}} \partial^2_\mu \hat{G}^{(1)}_\textbf{R} \ket{I^{(0)}_{\textbf{R}}}\\ & - \sum_\mu \bra{I'^{(0)}_{\textbf{R}}} \partial_\mu \hat{G}^{(1)}_\textbf{R} \ket{\partial_\mu I^{(0)}_{\textbf{R}}} = -\frac{1}{2} \sum_{\mu} \partial_\mu A^{(1)}_{I'I\mu,\textbf{R}}.
\end{split}
\end{equation}
The first line of Eq.~(\ref{Dpot2}) follows from the expansion of the commutator with the nuclear kinetic energy in Eq.~(\ref{Dpot1}), and in the second line I apply an identity proved in Appendix~\ref{appendixb} together with the definition of the first-order Berry connection in Eq.~(\ref{Vpot1}).

Collecting the results from the perturbative treatment of the non-adiabatic coupling matrices, Eqs.~(\ref{Vpot1}) and (\ref{Dpot2}), and the adiabatic potential energy matrix, Eq.~(\ref{expansion}), gives an expansion of the Born-Oppenheimer molecular Hamiltonian, Eq.(\ref{BOHam}), in the spin-orbit interaction:  
\begin{equation}\label{BOHamexp}
\begin{split}
    \hat{H}^{BO}_{I'I} = &\delta_{I'I}\hat{T}_{\textbf{R}} 
    -i A^{(1)}_{I'I\mu,\textbf{R}} \; \hat{p}_{\mu} - \frac{1}{2} \sum_{\mu} \partial_\mu A^{(1)}_{I'I\mu,\textbf{R}} \\ &+\delta_{I'I} E^{(0)}_{I,\textbf{R}} + E^{(1)}_{I'I,\textbf{R}} + E^{(2)}_{I'I,\textbf{R}}.
\end{split}
\end{equation}
Uniting the Berry connection terms with the nuclear kinetic energy gives a covariant form of this Hamiltonian:  
\begin{equation}\label{BOHamexp2}
\begin{split}
    \hat{H}^{BO}_{I'I} = &- \frac{1}{2} \sum_{\mu} \left( \partial_\mu + A^{(1)}_{I'I\mu,\textbf{R}} \right)^2 \\ &+ \delta_{I'I}  E^{(0)}_{I,\textbf{R}} + E^{(1)}_{I'I,\textbf{R}} + E^{(2)}_{I'I,\textbf{R}}.
\end{split}
\end{equation}
Eq.~(\ref{BOHamexp2}) and Eq.~(\ref{BOHamexp}) agree to first-order in the Berry connection upon expansion of the kinetic energy term. I use Eq.~(\ref{Vpot4}) and Eq.~(\ref{expansion}) in Eq.~(\ref{BOHamexp2}) to derive the final form of the new molecular spin Hamiltonian in terms of spin operators, $\hat{S}_{\alpha}$, acting in the ground-state multiplet space:
\begin{equation}\label{molecHspin}
\begin{split}
    \hat{H}^{\text{spin}} =& - \frac{1}{2} \sum_{\mu} \left( \partial_\mu + i \hat{S}_{\alpha} a^{(1)}_{\alpha\mu,\textbf{R}} \right)^2 \\ &+ E^{(0)}_{\textbf{R}} + \mu_B B_{\alpha} g^{(2)}_{\alpha\beta,\textbf{R}} \hat{S}_{\beta} + \hat{S}_\alpha D^{(2)}_{\alpha\beta,\textbf{R}} \hat{S}_\beta.
\end{split}
\end{equation}
Eq.~(\ref{molecHspin}), the extended molecular spin Hamiltonian, is a main result of this paper. It contains a novel, spin-vibronic vector potential that derives from the first-order Berry connection, as well as the traditional spin Hamiltonian terms that derive from the adiabatic potential energy matrix. The spin-vibronic vector potential gives rise to an effective magnetic field on the ground-state multiplet, and I explore its effects on the spin-vibrational dynamics in the rest of this paper. The spin-orbit interaction makes a leading first-order contribution to the vector potential, unlike the adiabatic molecular Zeeman and zero-field splitting terms, which are of second order in the spin-orbit and orbital Zeeman interactions. I expect that the spin-vibronic vector potential dominates the spin-vibrational dynamics in weak external magnetic fields, and I demonstrate its fundamental role for the spin relaxation dynamics of a prototypical molecular qubit. The external magnetic field also makes a contribution to the Berry connection: at first order the contribution to the Berry connection is diagonal in the ground-state multiplet space and does not couple to the spin-vibrational dynamics. 

\subsection{Gauge transformation}\label{gauge}
I showed in Sec.~\ref{molspinHam} that the spin-vibrational dynamics in the ground-state multiplet is governed by the molecular spin Hamiltonian in Eq.~(\ref{molecHspin}), which contains a novel, spin-vibronic vector potential together with the traditional spin Hamiltonian interactions. In Sec.~\ref{gauge}, I transform the representation of the Hamiltonian to bring the molecular spin Hamiltonian to a symmetrized form that is convenient for application in time-dependent perturbation theory.

Gauge transformations\cite{mead1992geometric,shapere1989geometric} are nuclear configuration-dependent single-valued unitary transformations of the electronic multiplet wavefunctions:
\begin{equation}\label{utranform}
     \ket{\tilde{I}_{\textbf{R}}} = \ket{I_{\textbf{R}}} U_{I'I,\textbf{R}} = \sum_{M'} \ket{iSM',\textbf{R}} U_{M'M,\textbf{R}},
\end{equation}
which preserve the dynamics generated by the Born-Oppenheimer molecular Hamiltonian, Eq.~(\ref{BOHam}). As a result of the gauge transformation in Eq.~(\ref{utranform}), the spin-vibronic vector potential, Eq.~(\ref{Berryconn}), transforms as a non-Abelian gauge potential,\cite{mead1992geometric,shapere1989geometric}
\begin{equation}\label{vtransform}
     \tilde{A}_{I'I\mu,\textbf{R}} = U^{\dagger}_{I'I'',\textbf{R}} A_{I''I'''\mu,\textbf{R}} U_{I'''I,\textbf{R}} + U^{\dagger}_{I'I'',\textbf{R}} \partial_\mu U_{I''I,\textbf{R}},
\end{equation}
and the Born-Oppenheimer Hamiltonian undergoes a unitary transformation,
\begin{equation}\label{htransform}
     \tilde{H}^{BO}_{I'I,\textbf{R}} = U^{\dagger}_{I'I'',\textbf{R}} H^{BO}_{I''I''',\textbf{R}} U_{I'''I,\textbf{R}}.
\end{equation}
Physical observables are independent of the specific choice of electronic representation; they are gauge-covariant quantities, which transform upon gauge transformations like the Hamiltonian in Eq.~(\ref{htransform}). For instance, the gauge-covariant nuclear momentum, called kinematic momentum $\hat{\Pi}_{I'I\mu}$, differs from the canonical momentum $\hat{p}_\mu$ by the vector potential:
\begin{equation}\label{kinematicp}
     \hat{\Pi}_{I'I\mu} = \delta_{I'I} \hat{p}_\mu -i A_{I'I\mu,\textbf{R}},
\end{equation}
and neither $\hat{p}_\mu$ nor $A_{I'I\mu,\textbf{R}}$ are separately gauge-covariant. The vector potential in Eq.~(\ref{kinematicp}) results in the non-commutativity of the kinematic momenta:
\begin{equation}\label{kinematicpcomm}
     \left[ \hat{\boldsymbol{\Pi}}_{\mu}, \hat{\boldsymbol{\Pi}}_{\nu} \right]_{I'I} = -F_{I'I\mu\nu,\textbf{R}},
\end{equation}
unlike the well-known commutation relation of the canonical momenta, $[\hat{p}_\mu,\hat{p}_\nu]=\delta_{\mu\nu}$. In Eq.~(\ref{kinematicpcomm}), bold symbols denote matrices in the ground-state multiplet space, and $F_{I'I\mu\nu,\textbf{R}}$ is the field tensor (also called Berry curvature\cite{berry1984quantal}):
\begin{equation}\label{fieldstrength}
     F_{I'I\mu\nu,\textbf{R}} = \partial_\mu A_{I'I\nu,\textbf{R}} - \partial_\nu A_{I'I\mu,\textbf{R}} + \left[ \textbf{A}_{\mu,\textbf{R}} , \textbf{A}_{\nu,\textbf{R}} \right]_{I'I},
\end{equation}
which is a gauge-covariant measure of the strength of the induced effective magnetic field on the ground-state manifold. The gauge-covariance of the field tensor requires that $U_{I'I\mathbf{R}}$ is both unitary and single-valued. Expansion of the spin-vibronic vector potential to first order in the spin-orbit interaction as in Sec.~\ref{molspinHam} gives the first-order spin-vibronic magnetic field tensor:
\begin{equation}\label{fieldstrength1}
     F^{(1)}_{I'I\mu\nu,\textbf{R}} = \partial_\mu A^{(1)}_{I'I\nu,\textbf{R}} - \partial_\nu A^{(1)}_{I'I\mu,\textbf{R}} = i S_{M'M,\alpha} f^{(1)}_{\alpha \mu\nu,\textbf{R}},
\end{equation}
with
\begin{equation}\label{fieldstrength2}
     f^{(1)}_{\alpha \mu\nu,\textbf{R}} = \partial_\mu a^{(1)}_{\alpha\nu,\textbf{R}} - \partial_\nu a^{(1)}_{\alpha\mu,\textbf{R}} = \partial_\mu \wedge a^{(1)}_{\alpha\nu,\textbf{R}}.
\end{equation}
In Eq.~(\ref{fieldstrength2}), $\wedge$ denotes the antisymmetric product, the generalization of the cross-product $\cross$ to multiple dimensions. With the first-order spin-vibronic vector potential, the kinematic momentum becomes:
\begin{equation}\label{kinmomentum1}
    \hat{\Pi}_{M'M\mu} = \delta_{M'M} \hat{p}_\mu + S_{M'M,\alpha} a^{(1)}_{\alpha\mu\nu,\textbf{R}}.
\end{equation}
The nuclear dynamics depends only on gauge-covariant quantities, and I derive the Heisenberg equation of motion for the kinematic momentum Eq.~(\ref{kinmomentum1}) with the Hamiltonian Eq.~(\ref{molecHspin}) to leading order in the spin-orbit interaction:
\begin{equation}\label{HeisenbergEOM}
\begin{split}
    \frac{d^2 \hat{R}_{\mu}}{dt^2}& =  \hat{S}_{\alpha} \frac{1}{2} \left\{ \frac{d \hat{R}_{\nu}}{dt} f^{(1)}_{\alpha \nu\mu,\textbf{R}} - f^{(1)}_{\alpha \mu\nu,\textbf{R}} \frac{d \hat{R}_{\nu}}{dt} \right\} \\ &- \partial_\mu E^{(0)}_{\textbf{R}} - \mu_B B_{\alpha} \partial_\mu g^{(2)}_{\alpha\beta,\textbf{R}} \hat{S}_{\beta} - \hat{S}_\alpha \partial_\mu D^{(2)}_{\alpha\beta,\textbf{R}} \hat{S}_\beta.
\end{split}
\end{equation}
The derivation of Eq.~(\ref{HeisenbergEOM}) uses the equation of motion for the nuclear position $\delta_{M'M} d \hat{R}_\mu / dt = \hat{\Pi}_{M'M\mu}$ with the velocity operator $d \hat{R}_\mu / dt$. The first term on the right-hand side of Eq.~(\ref{HeisenbergEOM}) is a quantum Lorentz force, the quantum equivalent to the classical Lorentz force\cite{jackson1999classical} $\mathbf{F}_L = \mathbf{v} \cross \mathbf{B}$ for a unit-charge particle moving in three dimensions in a magnetic field $\mathbf{B}$, where $\mathbf{B}$ relates to the electromagnetic tensor $F_{ij}$ as $\epsilon_{ijk} B_k = F_{ji}$ with the completely antisymmetric tensor $\epsilon_{ijk}$. Because of the non-Abelian algebraic structure of the spin-vibronic magnetic field tensor, there is a separate magnetic field for each generator of the associated non-Abelian group SO(3) or SU(2). The rest of the terms on the right-hand side are the forces that originate from the adiabatic potential energy matrix.

I use the freedom of gauge transformation to explicitly derive the spin-vibronic magnetic field tensor contribution to the spin Hamiltonian in Eq.~(\ref{molecHspin}). With the hindsight of the harmonic approximation, I first expand the spin-vibronic vector potential to linear order in the deviations $u_\mu$ from a fixed nuclear configuration $\textbf{R}_0$, which is equivalent to the multipole expansion of the electromagnetic vector potential\cite{sakurai2014modern}:
\begin{equation}\label{multipole}
    A^{(1)}_{I'I\mu,\textbf{R}} = A^{(1)}_{I'I\mu,\textbf{R}_0} + \partial_\nu A^{(1)}_{I'I\mu,\textbf{R}_0} u_\nu.
\end{equation}
I construct a single-valued unitary gauge transformation using the generator approach $U_{I'I,\textbf{R}} = e^{\Lambda_{I'I,\textbf{R}}}$ with the anti-Hermitian generator function:
\begin{equation}\label{gaugegen}
    \Lambda_{I'I,\textbf{R}} = A^{(1)}_{I'I\mu,\textbf{R}_0} u_\nu + \frac{1}{2} \partial_\nu A^{(1)}_{I'I\mu,\textbf{R}_0} u_\nu u_\mu,
\end{equation}
that transforms the molecular spin Hamiltonian to a symmetric gauge. I use this gauge transformation in Eq.~(\ref{vtransform}) to obtain the first-order, symmetrized spin-vibronic vector potential:
\begin{equation}\label{symvpot}
    \tilde{A}^{(1)}_{I'I\mu,\textbf{R}} = \frac{1}{2} \left( \partial_\nu A^{(1)}_{I'I\mu,\textbf{R}_0} u_\nu - \partial_\mu A^{(1)}_{I'I\nu,\textbf{R}_0} u_\nu \right).
\end{equation}
Gauge transforming Eq.~(\ref{molecHspin}) together with the vector potential in Eq.~(\ref{symvpot}) gives to leading order in the spin-orbit interaction the symmetrized molecular spin Hamiltonian:
\begin{equation}\label{molecHspin2}
\begin{split}
        \hat{H}^{\text{spin}} =& \hat{H}^{(0)} + \frac{1}{4} f^{(1)}_{\alpha\nu\mu,\textbf{R}_0} \hat{S}_{\alpha} \left( u_\nu \wedge \hat{p}_\mu \right) \\ &+ \mu_B B_{\alpha} g^{(2)}_{\alpha\beta,\textbf{R}} \hat{S}_{\beta} + \hat{S}_\alpha D^{(2)}_{\alpha\beta,\textbf{R}} \hat{S}_\beta,
\end{split}
\end{equation}
with
\begin{align*}\label{H0}
    \hat{H}^{(0)} = - \frac{1}{2} \sum_{\mu} \partial^2_\mu + E^{(0)}_{\textbf{R}} 
\end{align*}
and summation over repeated indeces. The significance of the spin-vibronic vector potential is particularly revealing in the symmetric gauge where it gives rise to a spin-vibrational orbit interaction, second term in Eq.~(\ref{molecHspin2}), which involves the effective magnetic field $f^{(1)}_{\alpha\nu\mu,\textbf{R}_0} \hat{l}_{\nu\mu}$ induced by the vibrational angular motion with vibrational angular momentum $\hat{l}_{\nu\mu} = u_\nu \wedge \hat{p}_\mu$. The spin-vibrational orbit interaction is analogous to the spin-electronic orbit interaction (SOI) and it is similarly symmetric upon both time-reversal and parity transformations. Note that the time-reversal invariance requires that the first-order magnetic field tensor $f^{(1)}_{\alpha\nu\mu,\textbf{R}}$ is real. This implies that in the absence of an external magnetic field, the spin-vibrational orbit interaction preserves Kramers' theorem\cite{sakurai2014modern} and guarantees that half-integer spin systems have doubly degenerate electronic states.

\subsection{Harmonic approximation}\label{harmonic}
I carried out a multipole expansion of the spin-vibronic vector potential followed by a gauge transformation to a symmetric gauge in Sec.~\ref{gauge} to derive the spin-vibrational orbit interaction in the molecular spin Hamiltonian Eq.~(\ref{molecHspin2}). In Sec.~\ref{harmonic}, I derive a harmonic approximation to the molecular spin Hamiltonian in Eq.~(\ref{molecHspin2}), which I use to obtain an expression for the spin relaxation time in Sec.~\ref{spinrate} using time-dependent perturbation theory. I start with the case of harmonic vibrational dynamics of molecules in Sec.~\ref{molvib}, which I extend to molecular crystals in Sec.~\ref{crystvib}. 

\subsubsection{Vibrations of molecules}\label{molvib}

The harmonic approximation to the vibrational dynamics of molecules\cite{wilson1980molecular} is based on the second-order expansion of the ground-state potential energy surface $E^{(0)}_{\textbf{R}}$ in $\hat{H}^{(0)}$ around the equilibrium nuclear geometry $\textbf{R}_0$:
\begin{equation}\label{Hharm}
\begin{split}
    \hat{H}^{\text{vib}} &= - \frac{1}{2} \sum_{\mu} \partial^2_\mu + \frac{1}{2} \partial_\mu \partial_\nu E^{(0)}_{\textbf{R}_0} u_\mu u_\nu \\ &= \frac{1}{4} \sum_i \omega_i \left( P_i^2 + Q_i^2 \right)
\end{split}
\end{equation}
I introduce in Eq.~(\ref{Hharm}) the dimensionless normal mode coordinates $Q_i$ and momenta $P_i$ of vibrations with frequencies $\omega_i$. The mass-weighted Cartesian displacements and momenta relate to the normal mode coordinates and momenta by the linear transformation:
\begin{align*}\label{Normalmodes}
    u_{\mu} &= \sum_{i} C_{\mu, i} \sqrt{\frac{1}{2\omega_i}} Q_i \;, \; Q_i = b_{i}^{\dagger} + b_i\\
    \hat{p}_\mu &= \sum_{i} C_{\mu, i} \sqrt{\frac{\omega_i}{2}} P_i \;, \; P_i = i \left( b_{i}^{\dagger} - b_i \right)
\end{align*}
with $C_{\mu, i}$ the eigenvectors of the mass-weighted Hessian matrix. The equations also give the expressions for the quantization of $Q_i$ and $P_i$ in terms of the ladder operators $b_i$ and $b^\dagger_i$. The second-order expansion of the spin-dependent terms in the Hamiltonian Eq.~(\ref{molecHspin2}) gives: (i) the traditional static spin Hamiltonian at the equilibrium nuclear configuration, which I call the reference spin Hamiltonian:
\begin{equation}\label{Hamspin}
\begin{split}
    \hat{H}^{\text{spin-ref}} = \mu_B B_{\alpha} g^{(2)}_{\alpha\beta,\textbf{R}_0} \hat{S}_{\beta} + \hat{S}_\alpha D^{(2)}_{\alpha\beta,\textbf{R}_0} \hat{S}_\beta;
\end{split}
\end{equation}
(ii) one-phonon (1P) spin-vibrational coupling Hamiltonian with adiabatic (AD) origin, deriving from the adiabatic potential energy matrix in Eq.~(\ref{molecHspin2}):
\begin{equation}\label{1phononCart}
\begin{split}
    \hat{H}^{\text{1P-AD}} =& \mu_B \partial_\nu g^{(2)}_{\alpha\beta,\textbf{R}_0} B_{\alpha} \hat{S}_{\beta} u_\nu + \partial_\nu D^{(2)}_{\alpha\beta,\textbf{R}_0} \hat{S}_\alpha \hat{S}_\beta u_\nu;
\end{split}
\end{equation}
(iii) two-phonon (2P) spin-vibrational coupling Hamiltonian with non-adiabatic (NA) origin, deriving from the spin-vibrational orbit interaction in Eq.~(\ref{molecHspin2}), as well as the adiabatic two-phonon spin-vibrational coupling Hamiltonian:
\begin{equation}\label{2phononCart}
\begin{split}
    \hat{H}^{\text{2P-NA}} = &\frac{1}{4} f^{(1)}_{\alpha\nu\mu,\textbf{R}_0} \hat{S}_{\alpha} \left( u_\nu \wedge \hat{p}_\mu \right), \\
    \hat{H}^{\text{2P-AD}} = &\frac{\mu_B}{2} \partial_\nu \partial_\mu  g^{(2)}_{\alpha\beta,\textbf{R}_0} B_{\alpha} \hat{S}_{\beta} u_\nu u_\mu + \\ &\frac{1}{2} \partial_\nu \partial_\mu D^{(2)}_{\alpha\beta,\textbf{R}_0} \hat{S}_\alpha \hat{S}_\beta u_\nu u_\mu.
\end{split}
\end{equation}
Transforming the spin-vibrational coupling Hamiltonians in Eqs.~(\ref{1phononCart}) and (\ref{2phononCart}) to normal mode representation gives: (i) the one-phonon spin-vibrational interactions:
\begin{equation}\label{DirectPert}
\begin{split}
    \hat{H}^{\text{1P-AD}} = &\mu_B \partial_i g^{(2)}_{\alpha\beta} B_{\alpha} \hat{S}^{\text{ref}}_{\beta} Q_i + \partial_i D^{(2)}_{\alpha\beta} \hat{S}^{\text{ref}}_\alpha \hat{S}^{\text{ref}}_\beta Q_i;
\end{split}
\end{equation}
(ii) the two-phonon spin-vibrational interactions:
\begin{equation}\label{RamanPert}
\begin{split}
    \hat{H}^{\text{2P-NA}} = &\frac{1}{4} f^{(1)}_{\alpha ji} \hat{S}^{\text{ref}}_{\alpha} \left( Q_j \wedge P_i \right), \\ 
    \hat{H}^{\text{2P-AD}} = &\frac{\mu_B}{2} \partial_i \partial_j  g^{(2)}_{\alpha\beta} B_{\alpha} \hat{S}^{\text{ref}}_{\beta} Q_i Q_j + \\ &\frac{1}{2} \partial_i \partial_j D^{(2)}_{\alpha\beta} \hat{S}^{\text{ref}}_\alpha \hat{S}^{\text{ref}}_\beta Q_i Q_j.
\end{split}
\end{equation}
In Eqs.~(\ref{DirectPert}) and (\ref{RamanPert}), indeces $i$ and $j$ refer to the dimensionless normal modes, and $f_{\alpha ji} = \omega_i \partial_j a_{\alpha i} - \omega_j \partial_i a_{\alpha j}$.

\subsubsection{Vibrations of molecular crystals}\label{crystvib}
The generalization of the harmonic approximation in Sec.~\ref{molvib} to molecular crystals requires a straightforward change of notation to account for the wavevector $\mathbf{q}$ dependence of the normal mode coordinates $Q_{i\mathbf{q}}$, momenta $P_{i\mathbf{q}}$, and frequencies $\omega_{i\mathbf{q}}$:\cite{ashcroft2022solid}
\begin{equation}\label{Hcrystal}
    \hat{H}^{\text{vib}} = \frac{1}{4} \sum_{i,\mathbf{q}} \omega_{i\mathbf{q}} \left( P_{i\mathbf{q}}^2 + Q_{i\mathbf{q}}^2 \right).
\end{equation}
The sum in Eq.~(\ref{Hcrystal}) goes over all normal mode branches $i$ and all values of the wavevector $\mathbf{q}$ in the first Brillouin zone. The mass-weighted Cartesian displacements $u_{\mu\mathbf{L}}$ and momenta $\hat{p}_{\mu\mathbf{L}}$ of atomic coordinate $\mu$ in the cell centered at $\mathbf{L}$ in terms of the normal mode coordinates and momenta become:
\begin{align*}
    u_{\mu\mathbf{L}} &= \sum_{i,\mathbf{q}} C_{\mu, i\mathbf{q}} \sqrt{\frac{1}{2N\omega_{i\mathbf{q}}}} e^{i \mathbf{q} \cdot \mathbf{L}} \: Q_{i\mathbf{q}} \;, \; Q_{i\mathbf{q}} = b_{i\mathbf{q}}^{\dagger} + b_{i-\mathbf{q}} \\
    \hat{p}_{\mu\mathbf{L}} &= \sum_{i,\mathbf{q}} C_{\mu, i\mathbf{q}} \sqrt{\frac{ \omega_{i\mathbf{q}}}{2N}} e^{i \mathbf{q} \cdot \mathbf{L}} \: P_{i\mathbf{q}} \;, \; P_{i\mathbf{q}} = i \left( b_{i\mathbf{q}}^{\dagger} - b_{i-\mathbf{q}} \right)
\end{align*}
with $C_{\mu, i\mathbf{q}}$ the eigenvectors of the mass-weighted dynamical matrix. The equations give also the quantization of the crystal vibrations in terms of the ladder operators $b_{i\mathbf{q}}$ and $b^{\dagger}_{i\mathbf{q}}$. The spin relaxation rate model in Sec.~\ref{Ramanrate} assumes that the paramagnetic molecular center, a paramagnetic molecule in a crystal environment of diamagnetic molecular counterparts, is localized in the unit cell at the origin $\mathbf{L}=0$ and relies only on the dynamics of the optical vibrational modes, for which I adopt an approximation that neglects the dependence of the optical mode eigenvectors $C_{\mu, i\mathbf{q}}$ on the wavevector $\mathbf{q}$. Within this approximation, the paramagnetic molecular center displacements and momenta in terms of the optical normal modes are given by:
\begin{equation}\label{opticalmodes}
\begin{split}
    u_{\mu} &= \sum_{i,\mathbf{q}} \sqrt{\frac{1}{2N\omega_i}} \: C_{\mu, i} \: Q_{i\mathbf{q}}, \\
    \hat{p}_{\mu} &= \sum_{i,\mathbf{q}} \sqrt{\frac{\omega_i}{2N}} \: C_{\mu, i} \: P_{i\mathbf{q}}.
\end{split}
\end{equation}
Substituting Eq.~(\ref{opticalmodes}) in Eq.~(\ref{2phononCart}) gives the expression for the two-phonon spin-vibrational coupling Hamiltonian of the paramagnetic molecular center,
\begin{equation}\label{RamanPertCrystal}
\begin{split}
    \hat{H}^{\text{2P-NA}} = &\frac{1}{N} \sum_{i\mathbf{q},j\mathbf{k}}  \frac{1}{4} f^{(1)}_{\alpha ji} \hat{S}^{\text{ref}}_{\alpha} \left( Q_{j\mathbf{k}} \wedge P_{i\mathbf{q}} \right) \\
    \hat{H}^{\text{2P-AD}} = &\frac{1}{N} \sum_{i\mathbf{q},j\mathbf{k}} \frac{\mu_B}{2} \partial_i \partial_j  g^{(2)}_{\alpha\beta} B_{\alpha} \hat{S}^{\text{ref}}_{\beta} Q_{j\mathbf{k}} Q_{i\mathbf{q}} + \\
    &\frac{1}{N} \sum_{i\mathbf{q},j\mathbf{k}} \frac{1}{2} \partial_i \partial_j D^{(2)}_{\alpha\beta} \hat{S}^{\text{ref}}_\alpha \hat{S}^{\text{ref}}_\beta Q_{j\mathbf{k}} Q_{i\mathbf{q}}. 
\end{split}
\end{equation}
Eq.~(\ref{RamanPertCrystal}) is the extension of the isolated molecule two-phonon spin-vibrational interactions Eq.~(\ref{RamanPert}) to a single paramagnetic molecular spin center in a diamagnetic crystal environment. I apply Eq.~(\ref{RamanPertCrystal}) in Sec.~\ref{spinrate} to compute the two-phonon Raman process contribution to the spin relaxation time. 

\subsection{Spin relaxation rate model}\label{spinrate}

\subsubsection{Rate model assumptions}\label{ratemodel}
The rate model in this paper focuses on the simulation of the spin relaxation dynamics of $S=1/2$ molecular qubits near room temperature and at weak to moderate magnetic field intensities. The rate model is motivated by experimental measurements of the temperature and magnetic field dependence of the spin-lattice relaxation time ($T_1$ time) of molecular qubit crystals under diamagnetic dilution.\cite{follmer2020understanding,zadrozny2015millisecond,fataftah2019metal,yu2019concentrated,amdur2022chemical,kazmierczak2023t,kazmierczak2022illuminating,kazmierczak2021impact,du1996electron} The assumptions of the rate model are:
\begin{itemize}
  \item the model applies to single paramagnetic molecular centers which are embedded in the crystal environment of diamagnetic molecular homologs. 

  \item the model targets early-transition-metal-based molecular qubits, allowing the perturbative treatment of the spin-orbit interaction of Sec.~\ref{molspinHam}. It further focuses on $S=1/2$ molecular qubits, for which the zero-field interaction is unnecessary, leaving the spin-vibronic magnetic-field and molecular Zeeman interactions as the only spin-dependent contributions to the Hamiltonian in Eq.~(\ref{molecHspin2}).

  \item the model assumes that the spin relaxation dynamics near room temperature is determined by the two-phonon spin-vibrational interactions in Eq.(\ref{RamanPertCrystal}), which involve a broad range of vibrational modes in the relaxation dynamics by the virtual absorption-emission of phonons, the two-phonon Raman process.

  \item the model includes only the intra-molecular, optical vibrational modes in the approximation of Eq.~(\ref{opticalmodes}) as they are more efficient in modulating the intra-molecular spin-dependent interactions.

  \item the model approximates the density-of-states of an optical vibrational mode by a Gaussian function centered at the isolated molecule normal mode frequency. The width of the Gaussian function is treated as an empirical parameter. 
\end{itemize}
Subsections \ref{refstate} and \ref{Ramanrate} develop the rate model based on time-dependent perturbation theory,\cite{mattuck1960spin,van1940paramagnetic,orbach1961spin} the molecular spin Hamiltonian of Sec.~\ref{harmonic}, and the model assumptions.

\subsubsection{Reference state}\label{refstate}
The application of time-dependent perturbation theory to evaluate the two-phonon Raman process rate of the model of Sec.~\ref{ratemodel} requires a definition of the initial and final states of the relaxation dynamics.\cite{mattuck1960spin} I assume that as a result of thermal equilibration the initial and final states are the eigenvectors of the reference spin Hamiltonian Eq.~(\ref{Hamspin}) at the equilibrium nuclear configuration:
\begin{equation}\label{Hamspinref}
\begin{split}
    & \hat{U}^{\dagger} \hat{H}^{\text{spin-ref}} \hat{U} = E_M \\
    & \hat{S}^{\text{ref}}_{\alpha} = \hat{U}^{\dagger} \hat{S}_{\alpha} \hat{U}.
\end{split}
\end{equation}
This definition of reference states can be straightforwardly implemented in the molecular spin Hamiltonian by rotating the quantization axis of the reference spin operators $\hat{S}^{\text{ref}}_{\alpha}$ as written in the second line of Eq.~(\ref{Hamspinref}), where $\hat{U}$ is the unitary matrix diagonalizing the reference spin Hamiltonian. With this definition, the quantization axis of $\hat{S}^{\text{ref}}_{\alpha}$ closely follows the direction of the external magnetic field for $S=1/2$ paramagnetic molecular centers with a near isotropic g-tensor. 

\subsubsection{Two-phonon Raman process rate}\label{Ramanrate}
The rate $k_{b \rightarrow a}$ for the spin-flip transition from the initial state $b$ to the final state $a$ via the two-phonon Raman process, where a phonon in mode $j\mathbf{k}$ is absorbed and a phonon in mode $i\mathbf{q}$ is emitted, is given by time-dependent perturbation theory as:
\begin{equation}\label{Ramanrate1}
\begin{split}
    k_{b \rightarrow a} = \frac{2\pi}{\hbar} & \sum_{i\mathbf{q},n_{i\mathbf{q}}} \sum_{j\mathbf{k},n_{j\mathbf{k}}} \rho_{n_{i\mathbf{q}}} \rho_{n_{j\mathbf{k}}} \delta\left( \hbar\omega_{i\mathbf{q}}-\hbar\omega_{j\mathbf{k}}-\Delta_{ba} \right) \\ &\abs{\bra{a,n_{i\mathbf{q}}+1,n_{j\mathbf{k}}-1}\hat{H}^{\text{2P}}\ket{b,n_{i\mathbf{q}},n_{j\mathbf{k}}}}^2. 
\end{split}
\end{equation}
In Eq.~(\ref{Ramanrate1}), the sum ranges over all optical vibrational modes according to the rate model assumptions in Sec.~\ref{ratemodel}, their associated wavevectors and occupation numbers $n_{i\mathbf{q}}$. $\rho_{n_{i\mathbf{q}}}$ is the thermal weight of mode $i\mathbf{q}$, $\omega_{i\mathbf{q}}$ - the frequency of the mode, and $\Delta_{ba}$ - the energy gap between the initial and final states. $\delta$ is the Dirac delta function that imposes energy conservation. The virtual absorption-emission process in Eq.~(\ref{Ramanrate1}) allows optical modes with close frequencies to match energetically the small energy gap in the spin-flip transition, justifying the leading role of the two-phonon Raman process in the near-room-temperature spin relaxation dynamics. Evaluation of the two-phonon spin-vibrational Hamiltonian matrix elements using Eq.~(\ref{RamanPertCrystal}) gives: (i) for the non-adiabatic two-phonon coupling matrix elements
\begin{equation}\label{RamanNA}
\begin{split}
&\bra{a,n_{i\mathbf{q}}+1,n_{j\mathbf{k}}-1}\hat{H}^{\text{2P-NA}}\ket{b,n_{i\mathbf{q}},n_{j\mathbf{k}}} = \\ &= i f^{(1)}_{\alpha ji} S^{\text{ref}}_{\alpha, ab} \: \frac{1}{N} \sqrt{n_{j\mathbf{k}}} \sqrt{n_{i\mathbf{q}}+1} \\ &= H^{\text{2P-NA}}_{ji,ab} \frac{1}{N} \sqrt{n_{j\mathbf{k}}} \sqrt{n_{i\mathbf{q}}+1},
\end{split}
\end{equation}
with the definition of $H^{\text{2P-NA}}_{ji,ab}$ in the last line of Eq.~(\ref{RamanNA}); (ii) for the adiabatic two-phonon coupling matrix elements
\begin{equation}\label{RamanAD}
\begin{split}
&\bra{a,n_{i\mathbf{q}}+1,n_{j\mathbf{k}}-1}\hat{H}^{\text{2P-AD}}\ket{b,n_{i\mathbf{q}},n_{j\mathbf{k}}}= \\ &= \mu_B \partial_i \partial_j  g^{(2)}_{\alpha\beta} B_{\alpha} S^{\text{ref}}_{\beta, ab} \: \frac{1}{N}  \sqrt{n_{j\mathbf{k}}} \sqrt{n_{i\mathbf{q}}+1} \\ &= H^{\text{2P-AD}}_{ji,ab} \frac{1}{N} \sqrt{n_{j\mathbf{k}}} \sqrt{n_{i\mathbf{q}}+1}
\end{split}
\end{equation}
with the definition of $H^{\text{2P-AD}}_{ji,ab}$ in the last line of Eq.~(\ref{RamanAD}). I include only the g-tensor contribution in Eq.~(\ref{RamanAD}) because I consider only $S=1/2$ systems, for which the D-tensor term vanishes. With the model for the optical vibrational modes in Eq.~(\ref{opticalmodes}), the Hamiltonian matrix elements are independent of the mode wavevector, resulting in the simplified expression for the Raman process rate:  
\begin{equation}\label{Ramanrate2}
\begin{split}
    k_{b \rightarrow a}& = \frac{2\pi}{\hbar} \sum_{i,j} \abs{H^{\text{2P-NA}}_{ji,ab}+ H^{\text{2P-AD}}_{ji,ab}}^2 \\ &\frac{1}{N^2} \sum_{\mathbf{q},\mathbf{k}} \delta\left( \hbar\omega_{i\mathbf{q}}-\hbar\omega_{j\mathbf{k}}-\Delta_{ba} \right) (\bar{n}(\omega_{i\mathbf{q}}) + 1) \bar{n}(\omega_{j\mathbf{k}}).
\end{split}
\end{equation}
I carry out the sum over the thermal weight in Eq.~(\ref{Ramanrate2}) to obtain the average thermal occupation of the vibrational mode $\bar{n}(\omega_{i\mathbf{q}})$. Furthermore, I carry out the sum over the mode wavevectors in Eq.~(\ref{Ramanrate2}) using the vibrational mode density of states $g_i(\omega)=\frac{1}{N} \sum_{\mathbf{q}} \delta(\omega - \omega_{i\mathbf{q}})$ to obtain the final expression for the Raman process rate:  
\begin{equation}\label{Ramanrate3}
\begin{split}
    &k_{b \rightarrow a} = \frac{2\pi}{\hbar^2} \sum_{i,j} \abs{H^{\text{2P-NA}}_{ji,ab}+ H^{\text{2P-AD}}_{ji,ab}}^2 \\ &\int d\omega g_i(\omega-\omega_{ba}) g_j(\omega) (\bar{n}(\omega-\omega_{ba}) + 1) \bar{n}(\omega).
\end{split}
\end{equation}
The spin-lattice ($T_1$) relaxation time characterizes the timescale of thermal equilibration of the spin system as a result of the spin-vibrational interactions with the phonon bath. For the case of an $S=1/2$ spin system that interacts with a phonon bath of reciprocal temperature $\beta$ with rate constants $k_{b \rightarrow a}$ and $k_{a \rightarrow b}$ satisfying detailed balance, the $T_1$ time is given by:
\begin{equation}\label{T1rate}
\begin{split}
    \frac{1}{T_1} = k_{b \rightarrow a} + k_{a \rightarrow b} \approx 2k.
\end{split}
\end{equation}
The last equality in Eq.~(\ref{T1rate}) follows from the smallness of $\omega_{ba}$ compared to the vibrational mode frequencies, and $k$ is defined as the expression obtained from Eq.~(\ref{Ramanrate3}) for vanishing $\omega_{ba}$. Substitution of Eq.~(\ref{Ramanrate3}) in Eq.~(\ref{T1rate}) gives the expression for the contribution of the two-phonon Raman process to the $T_1$ time:
\begin{equation}\label{Ramanrelax}
\begin{split}
    \frac{1}{T_1} = &\frac{4\pi}{\hbar^2} \sum_{i,j} \abs{H^{\text{2P-NA}}_{ji,ab}+ H^{\text{2P-AD}}_{ji,ab}}^2 \\ &\int d\omega g_i(\omega) g_j(\omega) (\bar{n}(\omega) + 1) \bar{n}(\omega).
\end{split}
\end{equation}
The implementation of Eq.~(\ref{Ramanrelax}) requires the vibrational mode density-of-states, for which I adopt the Gaussian approximation from Sec.~\ref{ratemodel} with mode frequency $\omega_i$ and width $\sigma_i$. With the Gaussian approximation, I arrive at the final expression for the $T_1$ spin relaxation time of the rate model of Sec.~\ref{ratemodel}:   
\begin{equation}\label{Ramanrelax2}
\begin{split}
    \frac{1}{T_1} =& \frac{4\pi}{\hbar^2} \sum_{i,j} \abs{H^{\text{2P-NA}}_{ji,ab}+ H^{\text{2P-AD}}_{ji,ab}}^2 \\ &\frac{1}{\sqrt{2\pi(\sigma_i^2+\sigma_j^2)}} e^{-\frac{1}{2} \frac{(\omega_i - \omega_j)^2}{\sigma_i^2+\sigma_j^2}} \frac{e^{-\beta\hbar\omega_{ji}}}{\left(1-e^{-\beta\hbar\omega_{ji}}\right)^2},
\end{split}
\end{equation}
where $\omega_{ji} = \frac{ \sigma_j^2 \omega_i + \sigma_i^2 \omega_j }{\sigma_i^2 + \sigma_j^2}$ is the center of the product of the mode density-of-states Gaussians. To obtain Eq.~(\ref{Ramanrelax2}), I assumed that the thermal probabilities change slowly at the scale of the mode density-of-states, allowing their evaluation at the center of the Gaussian product.
Section \ref{dft} presents the computation of the spin-vibrational Hamiltonian matrix elements in Eq.~(\ref{Ramanrelax2}) based on density functional theory calculations on the isolated paramagnetic molecule. The only empirical parameters that enter in the implementation of Eq.~(\ref{Ramanrelax2}) are the widths of the mode density-of-states, for which I use a simple model with mode widths: $\sigma_i = 10 \text{cm}^{-1}$ for $\omega_i < 100 \text{cm}^{-1}$, $\sigma_i = 5 \text{cm}^{-1}$ for $\omega_i < 200 \text{cm}^{-1}$, and $\sigma_i = 1 \text{cm}^{-1}$ for $\omega_i > 200 \text{cm}^{-1}$. The \textit{ab initio} estimation of the mode density-of-states requires a computational model of the molecular crystal, which is outside of the scope of the present work. I present results in Sec.~\ref{results} for the two-phonon Raman contribution to the $T_1$ time as a function of temperature and magnetic field orientation for the prototypical molecular qubit Cu(II) porphyrin.

\section{Implementation}\label{dft}
I derived in Sec.~\ref{theory} an extended molecular spin Hamiltonian that contains a novel, non-adiabatic contribution from the spin-vibronic vector potential. In Sec.~\ref{dft}, I present the numerical evaluation of the spin-vibronic vector potential via state-of-the-art density functional theory. 

The starting point of the derivation is Eq.~(\ref{Vpot4}), which I evaluate using linear response unrestricted density functional theory. As derived in Refs. \cite{neese2007calculation,neese2001prediction}, the first-order perturbed Kohn-Sham electronic wavefunctions $\ket{\psi^{(1)}_{0, \alpha}}$ with respect to the spin-orbit interaction are given by:
\begin{equation}\label{dftwfn}
\begin{split} 
    \ket{\psi^{(1)}_{0, \alpha}} = \sum_{a,i,\sigma} U^{\text{SOI}}_{ai\sigma, \alpha} \ket{\psi^{a\sigma}_{i\sigma}},
\end{split}
\end{equation}
where I use the convention that $i,j,..$ denote occupied molecular orbitals (MOs) and $a,b,..$ - unoccupied (virtual) MOs in the ground-state determinant, and $\sigma$ is the MO spin projection. The sum in Eq.~(\ref{dftwfn}) ranges over all single excitations from the occupied to the unoccupied MOs of both spin projections. The coefficients $U^{\text{SOI}}_{ai\sigma}$ satisfy the coupled-perturbed self-consistent-field Kohn-Sham equations:
\begin{equation}\label{cpscf}
\begin{split} 
    \sum_{a,i} M^{\sigma \sigma}_{bj,ai} U^{\text{SOI}}_{ai\sigma, \alpha} = -h^{\text{SOI}}_{b\sigma j\sigma, \alpha},
\end{split}
\end{equation}
with the magnetic electronic Hessian $M^{\sigma' \sigma}_{bj,ai}$ given by:
\begin{equation}\label{magneticHessian}
\begin{split} 
    M^{\sigma' \sigma}_{bj,ai} =& (\epsilon_{a\sigma} - \epsilon_{i\sigma}) \delta_{bj,ai} \delta_{\sigma \sigma'} + \\ &c_{\text{HF}} (\bra{a_\sigma j_\sigma'}\ket{i_\sigma b_\sigma'} - \bra{a_\sigma b_\sigma'}\ket{i_\sigma j_\sigma'}).
\end{split}
\end{equation}
In Eq.~(\ref{magneticHessian}), $\epsilon_{i\sigma}$ are the Kohn-Sham orbital energies, $c_{\text{HF}}$ is the percentage of exact Hartree-Fock exchange in the density functional approximation, and $\bra{a_\sigma j_\sigma'}\ket{i_\sigma b_\sigma'}$ are the electron repulsion integrals. $h^{\text{SOI}}_{b\sigma j\sigma, \alpha}$ in Eq.~(\ref{cpscf}) are the matrix elements of the spin-orbit coupling operator as defined in Appendix~\ref{appendixa}.

\begin{table*}[t]
\begin{center}
\begin{tabular}{R{0.5cm} R{0.5cm} R{1cm} R{1cm} R{1.6cm} R{1.6cm} R{1.6cm} R{1.6cm} R{1.6cm} R{1.6cm}} 
 \hline\hline
 $i$ & $j$ & $\omega_i$ & $\omega_j$ & $|f_{ij}|$ PBE/SV & $|f_{ij}|$ PBE/TZ & $|f_{ij}|$ PBE0/SV & $|\partial_{ij} g|$ PBE/SV & $|\partial_{ij} g|$ PBE/TZ & $|\partial_{ij} g|$ PBE0/SV \\
 \hline
10 & 10 & 126.8 & 126.8 &  0.000 &  0.000 &  0.000 &  0.042 &  0.043 &  0.061 \\ 
10 & 11 & 126.8 & 126.8 & 13.500 & 16.189 &  7.995 &  0.018 &  0.018 &  0.027 \\ 
11 & 11 & 126.8 & 126.8 &  0.000 &  0.000 &  0.000 &  0.042 &  0.043 &  0.061 \\ 
13 & 13 & 204.3 & 204.3 &  0.000 &  0.000 &  0.000 &  0.029 &  0.030 &  0.039 \\ 
13 & 14 & 204.3 & 204.4 &  3.356 &  2.885 &  4.824 &  0.007 &  0.007 &  0.009 \\ 
14 & 14 & 204.4 & 204.4 &  0.000 &  0.000 &  0.000 &  0.029 &  0.030 &  0.039 \\  
12 & 16 & 196.5 & 233.6 &  0.000 &  0.000 &  0.000 &  0.071 &  0.076 &  0.117 \\ 
15 & 17 & 212.9 & 240.6 &  4.211 &  4.727 &  5.345 &  0.002 &  0.002 &  0.002 \\ 
18 & 19 & 288.9 & 288.9 & 13.961 & 14.213 & 15.205 &  0.003 &  0.003 &  0.012 \\ 
22 & 23 & 387.1 & 389.1 &  3.532 &  3.306 &  3.092 &  0.001 &  0.001 &  0.002 \\  
23 & 25 & 389.1 & 433.4 &  8.129 &  8.826 &  9.726 &  0.001 &  0.001 &  0.002 \\  
25 & 26 & 433.4 & 437.8 &  4.292 &  4.033 &  4.351 &  0.001 &  0.001 &  0.002 \\ 
28 & 29 & 454.3 & 454.3 &  7.686 &  8.606 &  2.158 &  0.000 &  0.000 &  0.000 \\   
 \hline\hline
\end{tabular}
\end{center}
\caption{Two-phonon spin-vibrational Hamiltonian matrix elements for the leading pairs of normal modes. $i$ and $j$ are the indeces of the modes, and $\omega_i$ and $\omega_j$ are the mode frequencies in $\text{cm}^{-1}$. $|f_{ij}|^2 = \sum_{\alpha} f^2_{\alpha ij}$ denotes the magnitude of the spin-vibronic magnetic field tensor for modes $i$ and $j$ in $10^{-3} \text{cm}^{-1}$. $|\partial_{ij} g|^2 = \mu_B \sum_{\alpha \ne \beta } \partial_i \partial_j g^2_{\alpha\beta ij}$ denotes the magnitude of the off-diagonal portion of the second derivative g-tensor for modes $i$ and $j$ in $10^{-3} \text{cm}^{-1}/\text{T}$. The matrix elements are computed by: PBE density functional and def2-SVP basis set (PBE/SV), PBE density functional and def2-TZVP basis set (PBE/TZ), and PBE0 density functional and def2-SVP basis set (PBE0/SV).}
\label{table:tab1}
\end{table*}

\begin{figure}
\includegraphics[width=0.85\linewidth]{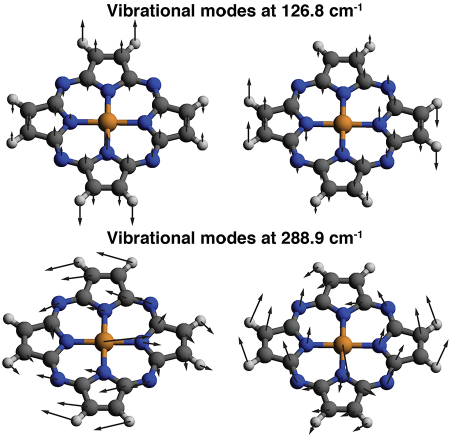}
\caption{\label{fig:fig1} Major normal modes contributing to the two-phonon spin relaxation time $T_1$ for the prototypical Cu(II) porphyrin $S=1/2$ molecular qubit. The four normal modes are doubly degenerate with normal mode frequencies $126.8 \text{cm}^{-1}$ (upper pair) and $288.9 \text{cm}^{-1}$ (lower pair). The mode at $126.8 \text{cm}^{-1}$ brings the molecule out of the symmetry plane, whereas the mode at $288.9 \text{cm}^{-1}$ develops within the plane of Cu(II) porphyrin. Arrows portray the relative magnitude and direction of the atomic displacements. Color code: Carbon is dark gray, Nitrogen - blue, Copper - orange, and Hydrogen - light gray.}
\end{figure}

With Eq.~(\ref{dftwfn}) the spin-vibronic vector potential in Eq.~(\ref{Vpot4}) becomes:
\begin{equation}\label{spinvibpot2}
\begin{split} 
    a^{(1)}_{\alpha \mu,\textbf{R}} &= -2 \sum_{a,i,\sigma} U^{\text{SOC}}_{ia\sigma, \alpha \mathbf{R}} \bra{a_{\sigma \mathbf{R}}} \ket{\partial_{\mu} i_{\sigma \mathbf{R}}}.
\end{split}
\end{equation}
With generalized-gradient density functional approximations, the coupled-perturbed Kohn-Sham equations have a non-iterative solution, and Eq.~(\ref{spinvibpot2}) gives:
\begin{equation}\label{spinvibpotgga}
\begin{split} 
    a^{(1)}_{\alpha \mu,\textbf{R}} &= 2 \sum_{a,i,\sigma} \frac{h^{\text{SOI}}_{i\sigma a\sigma, \alpha \mathbf{R}}}{\epsilon_{i\sigma, \mathbf{R}} - \epsilon_{a\sigma, \mathbf{R}}} \bra{a_{\sigma \mathbf{R}}} \ket{\partial_{\mu} i_{\sigma \mathbf{R}}},
\end{split}
\end{equation}
which is the result of Eq.~(\ref{spinvibpot2}) for electronic states represented as single Kohn-Sham Slater determinants.
To complete the evaluation of the spin-vibronic vector potential, I obtain the last term in the sum in Eq.~(\ref{spinvibpot2}) from the solution of the coupled-perturbed Kohn-Sham equations for the nuclear displacements\cite{bykov2015efficient,pople1979derivative}. These equations are conventionally solved when computing analytical second derivatives of the ground-state energy, and using the results of analytical gradient theory,\cite{bykov2015efficient,pople1979derivative} I write the nuclear derivatives of the Kohn-Sham MOs as:
\begin{equation}\label{angrad}
\begin{split} 
    \ket{\partial_{\mu} i_{\sigma \mathbf{R}}} = \sum_{p,i} U^{\text{G}}_{pi\sigma, \mu} \ket{p_{\sigma \mathbf{R}}} + \ket{\partial_{\mu} \tilde{i}_{\sigma \mathbf{R}}},
\end{split}
\end{equation}
where $U^{\text{G}}_{pi\sigma, \mu}$ are the solutions of the coupled-perturbed Kohn-Sham equations for the nuclear displacement $\mu$, and $\ket{\tilde{i}_{\sigma \mathbf{R}}}$ denotes MOs with frozen orbital coefficients, such that the nuclear derivative operates only on the atom-centered basis functions. Using Eq.~(\ref{angrad}) in Eq.~(\ref{spinvibpot2}), I arrive at the final formula for the evaluation of the spin-vibronic vector potential:
\begin{equation}\label{spinvibpotdft}
\begin{split} 
    a^{(1)}_{\alpha \mu,\textbf{R}} &= -2 \sum_{a,i,\sigma} U^{\text{SOC}}_{ia\sigma, \alpha \mathbf{R}} \left( U^{\text{G}}_{ai\sigma, \mu \mathbf{R}} + S_{ai\sigma, \mu \mathbf{R}} \right),
\end{split}
\end{equation}
where $S_{ai\sigma, \mu \mathbf{R}}$ is the right-hand derivative of the basis overlap matrix in the MO basis. The numerical evaluation of Eq.~(\ref{spinvibpotdft}) requires two coupled-perturbed Kohn-Sham response calculations: one response calculation for the spin-orbit interaction perturbation $U^{\text{SOI}}_{ia\sigma, \alpha \mathbf{R}}$, and a second response calculation for the nuclear displacement perturbations $U^{\text{G}}_{ai\sigma, \mu \mathbf{R}}$. I implement Eq.~(\ref{spinvibpotdft}) in an in-house version of the quantum chemistry software package ORCA\cite{neese2012orca}, for which I adapted already existing highly optimized algorithms for molecular g-tensor calculations\cite{neese2001prediction} and analytical second derivatives\cite{bykov2015efficient}. I applied in all calculations a widely-used mean-field approximation for the spin-orbit coupling operator\cite{neese2005efficient} that is already available in the ORCA suite.

\begin{figure}
\includegraphics[width=0.76\linewidth]{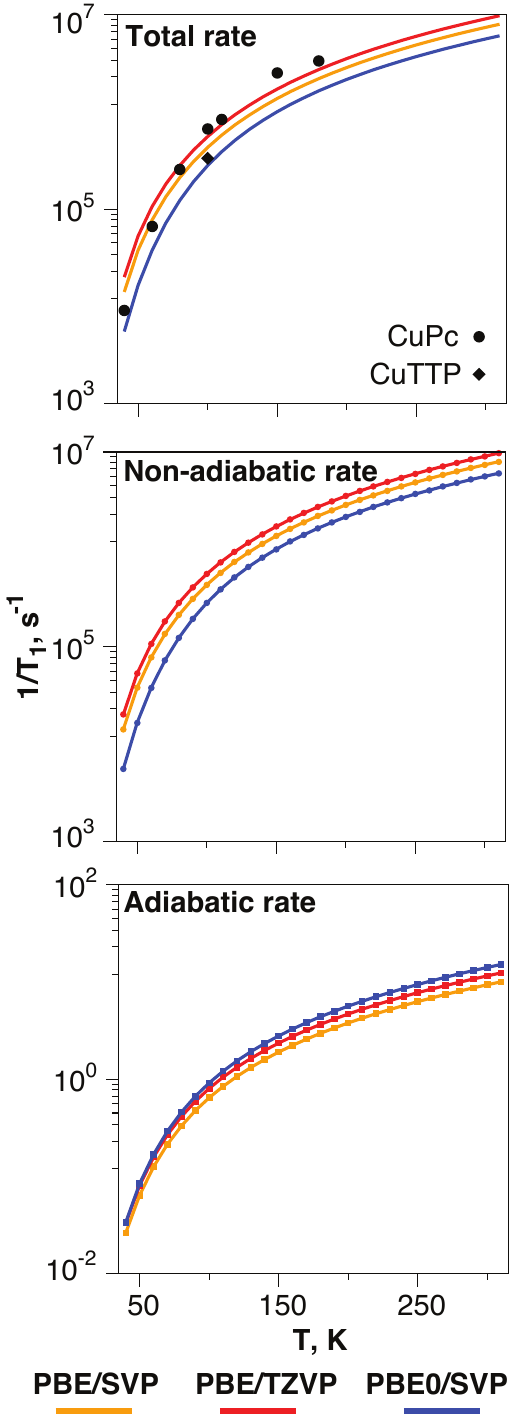}
\caption{\label{fig:fig2} Two-phonon spin relaxation rate $1/T_1$ as a function of temperature $T$ in K for Cu(II) porphyrin. Rate is in $\text{s}^{-1}$ and the y-axis is a logarithmic scale (base 10). The rate is uniformly averaged over the magnetic field orientation. Magnetic field intensity is $B=330\text{mT}$. Upper panel is the total rate, middle panel - the non-adiabatic contribution to the rate, and lower panel - the adiabatic contribution to the rate. Panels share the same temperature range. Black circles represent experimentally measured $T_1$ times for Cu(II)-phthalocyanine (CuPc) from Ref. \cite{follmer2020understanding}. Black diamond is an experimetally measured $T_1$ time for Cu(II) tetratolylporphyrin (CuTPP) from Ref. \cite{du1996electron}. Color code: orange - PBE/def2-SVP, red - PBE/def2-TZVP, and blue - PBE0/def2-SVP. Note the difference in scale of the adiabatic and the non-adiabatic rate.}
\end{figure}

\begin{table}[ht]
\begin{center}
\begin{tabular}{R{0.5cm} R{0.5cm} R{1cm} R{1cm} R{1.4cm} R{1.4cm} R{1cm}} 
 \hline\hline
 $i$ & $j$ & $\omega_i$ & $\omega_j$ & DoS & Th. Pr. &  $\frac{1}{T_1}$ \\
 \hline
10 & 11 &  126.8 &  126.8 &  0.056 &  2.588 &  1.758 \\
13 & 14 &  204.4 &  204.4 &  0.282 &  0.949 &  1.172 \\
18 & 19 &  288.9 &  289.0 &  0.282 &  0.438 &  5.385 \\
22 & 23 &  387.1 &  389.2 &  0.102 &  0.215 &  0.039 \\
28 & 29 &  454.3 &  454.3 &  0.282 &  0.141 &  0.035 \\
33 & 34 &  662.9 &  662.9 &  0.282 &  0.044 &  0.036 \\
 \hline\hline
\end{tabular}
\end{center}
\caption{Two-phonon spin relaxation rate contributions for room temperature $T=298 \text{K}$, magnetic field intensity $B=330 \text{mT}$, and magnetic field orientation $\theta = 90^{\text{o}}$ relative to the molecular axis. $i$ and $j$ are the indeces of the modes, and $\omega_i$ and $\omega_j$ are the mode frequencies in $\text{cm}^{-1}$. DoS denotes the mode density-of-states contribution to the rate using the Gaussian model in Sec.~\ref{spinrate}. Th. Pr. is the thermal probability contribution to the rate. $1/T_1$ is the two-phonon spin relaxation rate in MHz. The spin-vibronic matrix elements are calculated at PBE0/def2-SVP level of theory.}
\label{table:tab2}
\end{table}

The implementation of the spin-vibronic vector potential Eq.~(\ref{spinvibpotdft}) is consistent with the numerical evaluation of the g-tensor\cite{neese2001prediction} and the D-tensor\cite{neese2007calculation} in linear response unrestricted density functional theory. The molecular g-tensor, for instance, is expressed as:
\begin{equation}\label{gtensordft}
\begin{split} 
    g^{(2)}_{\alpha\beta,\textbf{R}} &= - \sum_{a,i,\sigma} U^{\text{SOI}}_{ia\sigma,\alpha\mathbf{R}} \; \text{Im}( L_{ai\sigma,\beta\mathbf{R}}),
\end{split}
\end{equation}
where $\text{Im}( L_{ai\sigma,\beta\mathbf{R}})$ is the imaginary part of the electronic orbital angular momentum operator in the MO basis. Both the spin-vibronic vector potential Eq.~(\ref{spinvibpotdft}) and the molecular g-tensor Eq.~(\ref{gtensordft}) expressions contain the self-consistent perturbation of the Kohn-Sham orbitals with respect to the spin-orbit interaction; the crucial difference is the second perturbing interaction: for the spin-vibronic vector potential, these are the nuclear displacement perturbations, whereas for the molecular g-tensor this is the orbital Zeeman interaction. This difference determines the different orders of the two effective interactions: the spin-vibronic vector potential is first order in the spin-orbit interaction, whereas the molecular g-tensor is first order in the spin-orbit interaction but also first order in the small orbital Zeeman interaction, resulting in an overall second order expression.

The matrix elements that enter in the calculation of the spin relaxation time Eq.~(\ref{RamanPertCrystal}) require geometric derivatives of both the spin-vibronic vector potential and the molecular g-tensor. I calculate these derivatives by numerical differentiation using a central difference approximation and normal mode displacements. I apply a dimensionless mode displacement (displacement in units of the zero-point amplitude) of 0.5. The calculation protocol starts with a gas-phase optimization of the molecular geometry using the selected basis set and density functional together with: resolution-of-identity approximation for the Coulomb portion of the Fock matrix,\cite{eichkorn1997auxiliary} fine grid for the exchange-correlation functional integration, very tight criteria for the self-consistent-field convergence, and tight criteria for the geometry convergence, all per ORCA 3.0.0 definitions.\cite{neese2012orca} A calculation of the analytical second derivatives follows using the same convergence criteria as applicable. In this work, I computed the molecular Hessian using the Perdew-Burke-Ernzerhof (PBE) density functional\cite{perdew1996generalized} and the Ahlrichs def2-TZVP basis set.\cite{schafer1992fully,schafer1994fully,weigend2005balanced} I calculated the spin-vibronic magnetic field tensor and the molecular g-tensor second geometrical derivatives using three density functional and basis set combinations: PBE functional and def2-SVP basis set, PBE functional and def2-TZVP basis set, and the hybrid PBE0 functional\cite{adamo1999toward,ernzerhof1999assessment} and def2-SVP basis set. The results of these calculations and the associated spin-lattice relaxation times for the Cu(II) porphyrin molecular qubit are presented in Sec.~\ref{results}. 

\begin{table}[ht]
\begin{center}
\begin{tabular}{R{0.5cm} R{0.5cm} R{1cm} R{1cm} R{1cm} R{1cm}} 
 \hline\hline
 $i$ & $j$ & $f_{jix}$ & $f_{jiy}$ & $f_{jiz}$ &  $\frac{1}{T_1}$ \\
 \hline
10 & 11 & 0.000 &  0.000 &  1.999 &  1.758 \\
13 & 14 & 0.000 & -0.003 & -1.206 &  1.172 \\
18 & 19 & 0.000 &  0.000 &  3.801 &  5.385 \\
22 & 23 & 0.000 &  0.000 & -0.773 &  0.039 \\
28 & 29 & 0.004 &  0.000 &  0.539 &  0.035 \\
33 & 34 & 0.000 &  0.000 &  0.971 &  0.036 \\
 \hline\hline
\end{tabular}
\end{center}
\caption{Spin-vibronic magnetic-field tensor components for the leading contributions to the two-phonon spin relaxation rate for room temperature $T=298 \text{K}$, magnetic field intensity $B=330 \text{mT}$, and magnetic field orientation $\theta = 90^{\text{o}}$ relative to the molecular symmetry axis. $i$ and $j$ are the indeces of the modes. $f_{jix}$ denotes the value of the spin-vibronic tensor $x$-component in $10^{-3} \text{cm}^{-1}$. $1/T_1$ is the two-phonon spin relaxation rate in MHz. The spin-vibronic matrix elements are calculated at PBE0/def2-SVP level of theory.}
\label{table:tab3}
\end{table}

\section{Results}\label{results}
I apply the new molecular spin Hamiltonian to the spin relaxation dynamics of Cu(II) porphyrin, a prototypical $S=1/2$ molecular qubit that represents the common core of a homologous series of Cu(II)-based molecular qubits. I present the results for the two-phonon spin relaxation time using the spin rate model of Sec.~\ref{spinrate}, and I calculate the two-phonon spin-vibrational Hamiltonian matrix elements according to the implementation in Sec.~\ref{dft} using three levels of density functional theory to investigate the density functional and basis set dependence of the computed matrix elements.

Table~\ref{table:tab1} presents the magnitudes of the spin-vibronic magnetic field tensor and the off-diagonal second derivative g-tensor for the leading pairs of normal modes. The striking result is that the spin-vibronic magnetic field matrix elements are two to three orders of magnitude larger than the derivative g-tensor matrix elements, demonstrating that the dominating spin-vibrational coupling mechanism in this class of molecular qubits is of non-adiabatic character. The results in Table~\ref{table:tab1} affirm that this conclusion is independent of both the basis set size and the density functional approximation. The basis set dependence of the matrix elements is mild, showing that computationally affordable basis sets of double zeta size can be employed to compute spin-vibronic coupling matrix elements. The amount of Hartree-Fock exchange in the density functional (the PBE0 density functional has $25\%$ exact exchange compared to $0\%$ for the PBE functional) leads to larger deviations in the magnitudes of the matrix elements, but the values are in quantitative agreement with one another, allowing the use of generalized gradient density functionals for initial investigations.

\begin{figure}
\includegraphics[width=0.76\linewidth]{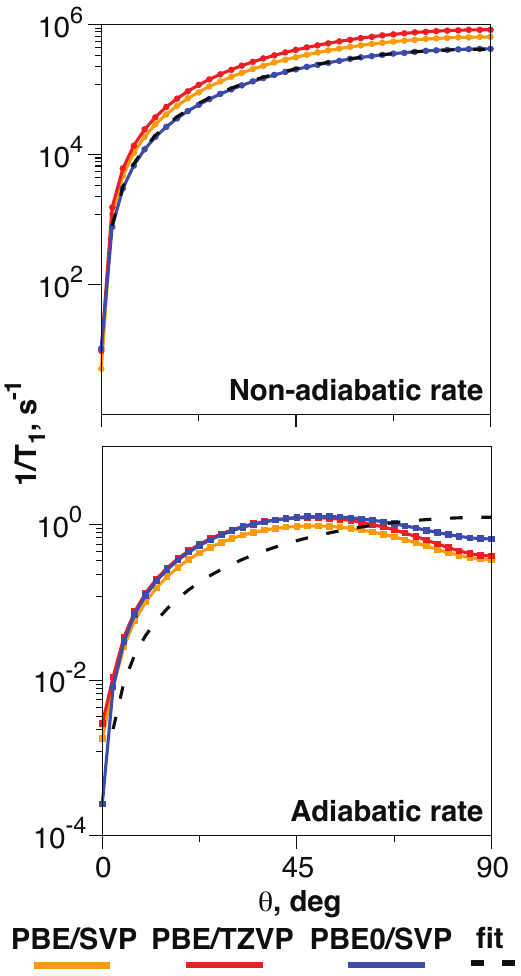}
\caption{\label{fig:fig3} Two-phonon spin relaxation rate $1/T_1$ as a function of the angle $\theta$ in degrees relative to the $C_4$ axis of Cu(II) porphyrin. Rate is in $\text{s}^{-1}$ and the y-axis is a logarithmic scale (base 10). The rate is uniformly averaged over the polar angle $\phi$ in the plane of the molecule. Temperature is 100K and magnetic field intensity is 330 mT. Upper panel is the the non-adiabatic contribution to the rate and lower panel - the adiabatic contribution to the rate. Panels share the same angle range. The non-adiabatic rate is practically equal to the total rate - note the difference in the scales of the two panels.  Color codes the level of theory: orange - PBE/def2-SVP, red - PBE/def2-TZVP, and blue - PBE0/def2-SVP. The dashed black line is a $A\text{sin}^2(\theta)$ fit to the PBE0/def2-SVP data. The fit to the non-adiabatic contribution coincides with the computed results.}
\end{figure}

The data in Table~\ref{table:tab1} further demonstrate that degenerate normal modes have a disproportionately large contribution to the spin-vibrational coupling mechanism in Cu(II) porphyrin. Figure~\ref{fig:fig1} depicts the two normal modes that are the major contributors to the spin-vibrational coupling in the qubit. The mode at $126.8 \text{cm}^{-1}$ brings the molecule out of the molecular plane, whereas the mode at $288.9 \text{cm}^{-1}$ develops in the molecular plane. Table~\ref{table:tab1} shows that the spin-vibronic matrix elements of the $288.9 \text{cm}^{-1}$ in-plane mode increase with the amount of Hartree-Fock exchange in the density functional and this mode becomes the major spin-vibrational coupling channel at PBE0/def2-SVP level of theory. The emerging physical picture from the results in Table~\ref{table:tab1} and Fig.~\ref{fig:fig1} is that nearly degenerate normal modes linearly superpose to give vibrational rotational modes, which strongly interact with the molecular spin via the spin-vibrational orbit interaction (see end of Sec.~\ref{gauge}). The spin-vibrational orbit interaction is of non-adiabatic origin and results from the non-Abelian Berry curvature on the ground-state electronic spin multiplet.

Figure~\ref{fig:fig2} presents the magnetic-field-orientation averaged two-phonon spin relaxation rate $1/T_1$, resulting from the coupling to the optical normal modes of the molecular qubit computed by the spin relaxation rate model of Sec.~\ref{spinrate}. The data in Figure~\ref{fig:fig2} confirm that the total two-phonon relaxation rate is determined by the non-adiabatic spin-vibrational orbit interaction and the adiabatic contribution to the relaxation rate is five orders of magnitude smaller. Similar to the spin-vibronic coupling, this conclusion holds true independently of the basis set size and the density functional approximation. The calculated rates are in excellent agreement with experimentally measured $T_1$ times for both a powder of Cu(II) phthalocyanine\cite{follmer2020understanding} and single crystals of Cu(II) tetratolylporphirin\cite{du1996electron}. The theory confirms that the leading spin relaxation mechanism in Cu(II)-based molecular qubits is the spin-vibrational orbit interaction.

The rates in Fig.~\ref{fig:fig2} display a characteristic activated dynamics with increasing temperature, which results from the successive thermal population of the doubly degenerate normal modes. Table~\ref{table:tab2} presents the major normal mode contributions to the two-phonon spin relaxation rate for $T=298 \text{K}$ and $B=330\text{mT}$. The data demonstrate that the doubly degenerate normal modes at $288.9 \text{cm}^{-1}$, $126.8 \text{cm}^{-1}$, and $204.4 \text{cm}^{-1}$ are the major channels for two-phonon spin relaxation in the system, and the three normal mode pairs fully determine the order of magnitude of the rate. This conclusion suggests an appealing route to control the spin relaxation rate of $S=1/2$ molecular qubits by decreasing the molecular symmetry via removal of symmetry axes greater than second order. 

Figure~\ref{fig:fig3} plots the two-phonon spin relaxation rate as a function of the angle $\theta$ between the molecular $C_4$ axis and the external magnetic field for $T=100 \text{K}$ and $B=330\text{mT}$. The relaxation rates are averaged over the angle $\phi$ that describes the rotation of the magnetic field vector around the $C_4$ axis. The results in Figure~\ref{fig:fig3} demonstrate the strong magnetic-field orientation dependence of the two-phonon relaxation rate with distinct functional dependence of the rate contributions on $\theta$: the non-adiabatic contribution, which entirely dominates the total rate, shows a clear $\text{sin}^2(\theta)$ dependence (dashed line), whereas the adiabatic contribution goes through a maximum at $45^{\text{o}}$ and is described poorly by the $\text{sin}^2(\theta)$ function. This orientational dependence of the two-phonon rate allows to distinguish theoretically and experimentally the non-adiabatic and the adiabatic mechanisms of spin relaxation. The computed $\text{sin}^2(\theta)$ dependence of the total rate is in complete agreement with the observed $\text{sin}^2(\theta)$ trend of the experimentally measured $T_1$ times\cite{du1996electron}, which presents an independent confirmation of the non-adiabatic spin relaxation mechanism. Furthermore, as a consequence of the orientational dependence, the two-phonon spin relaxation rate for in-plane orientation $\theta=90^{\text{o}}$ of the magnetic field is several orders of magnitude larger than the rate for perpendicular orientation $\theta=0^{\text{o}}$. This result leads to a greater than one $T_1$ anisotropy, defined as the ratio of the in-plane to the perpendicular relaxation rate, that similarly parallels experimental observations.

The dependence of the two-phonon rate on the orientation of the external magnetic field can be rationalized based on the data in Table~\ref{table:tab3}, which presents the components of the spin-vibronic magnetic field tensor for the leading normal mode pairs in the spin relaxation dynamics at $T=298K$ and $B=330\text{mT}$. Table~\ref{table:tab3} shows that the largest component of the spin-vibronic magnetic field tensor is along the $C_4$ axis of Cu(II) porphyrin and the in-plane components are smaller by several orders of magnitude. As a result, the induced effective magnetic field $f_{\alpha ij} \hat{l}_{ij}$ by the spin-vibrational orbit interaction is entirely directed along the $C_4$ axis of the molecule, and the most  efficient relaxation occurs when the molecular spin is oriented perpendicularly to the $C_4$ axis in the molecular plane. Provided that the molecular g-tensor at the equilibrium geometry is very nearly axial, the molecular spin closely follows the orientation of the external magnetic field, such that the in-plane molecular spin component varies as $\text{sin}(\theta)$ with $\theta$ the angle between the magnetic field vector and the $C_4$ axis. The relaxation rate depends on the squared modulus of the spin-vibrational orbit interaction, resulting in the observed $\text{sin}^2(\theta)$ dependence in Fig.~\ref{fig:fig3}.

\section{Conclusions}\label{conclusions}

I derive in the present paper an extended molecular spin Hamiltonian that takes into account both the traditional adiabatic spin-dependent interactions and includes a novel, non-adiabatic spin-vibrational interaction. The derivation employs the Born-Oppenheimer approximation to decouple the dynamics on different electronic spin manifolds, which induces a non-Abelian Berry connection on the ground-state electronic multiplet. The resulting non-Abelian Berry curvature is the spin-vibrational orbit interaction, which couples the vibrational angular momentum to the electronic spin in complete analogy to the spin-electronic orbit interaction. Thus, the vibrational angular motion induces an effective magnetic field in the spin dynamics on the electronic ground state and vice versa the electronic spin exerts a quantum Lorentz force on the nuclear motion. The spin-vibronic magnetic field tensor that quantifies the strength of coupling between the vibrational angular motion and the electronic spin is first order in the spin-orbit interaction and first order in the non-adiabatic interactions between the electronic multiplets. As such, this previously unappreciated interaction is expected to dominate the spin relaxation dynamics at weak magnetic field intensities and elevated temperatures.

I implement the computation of the matrix elements of the molecular spin Hamiltonian using linear response density functional theory, which allows the \textit{ab initio} prediction of the strength of the different interactions and the unbiased comparison between them. The density-functional implementation requires two different kinds of coupled-perturbed calculations for computing the response of the electronic wavefunctions to the spin-orbit interaction and to the nuclear displacements. I further develop a rate model to estimate the spin relaxation time using the molecular spin Hamiltonian computed for the isolated paramagnetic molecule. The rate model is specialized to isolated paramagnetic molecular centers in molecular crystals and to elevated temperatures and estimates the spin relaxation time via the two-phonon Raman process with the sole participation of the optical crystal vibrations. The only empirical parameters of the model are the widths of the mode density-of-states, for which I employ a Gaussian approximation.

I apply the molecular spin Hamiltonian together with the spin relaxation rate model to the prototypical $S=1/2$ molecular qubit Cu(II) porphyrin. The striking result is that the spin relaxation rate near room temperature is determined by the new spin-vibrational orbit interaction, and that the computed spin relaxation rate is in remarkable agreement with experimental measurements. This result is independent of both basis set size and density functional approximation. The physical picture that emerges for spin relaxation in Cu(II)-based molecular qubits is that the vibrational angular motion of nearly degenerate normal modes strongly couples to the electronic spin dynamics and is responsible for spin relaxation in the system. Furthermore, the spin relaxation time shows a strong dependence on the orientation of the magnetic field to the $C_4$ symmetry axis of the molecule and is distinct for the non-adiabatic and the adiabatic relaxation mechanisms, which allows to distinguish them both theoretically and experimentally.

The current work lays the foundation for a further investigation of a much broader set of $S=1/2$ qubits. Furthermore, the approach can be successfully applied to $S>1/2$ systems that include the emerging class of optically addressable molecular qubits. Taken together, the new molecular spin Hamiltonian formalism together with density functional theory is expected to provide a much-needed theoretical approach to simulate and understand spin relaxation dynamics in molecular qubits. 

\begin{acknowledgments}
I would like to acknowledge start-up funds from Indiana University, Bloomington and Tufts University.
\end{acknowledgments}

\section*{Data availability}
The data that support the findings of this study are available from the corresponding author upon reasonable request.

\begin{appendices}

\section{Spin structure via the Wigner-Eckart theorem}\label{appendixa}
In Appendix~\ref{appendixa}, I apply the Wigner-Eckart theorem to establish the spin-operator equivalents of the effective interactions in the molecular spin Hamiltonian. The spin-dependent relativistic interactions $\hat{H}^{\text{SOZ}}$, including the interaction with the weak external magnetic field, consists of: (i) the spin-orbit interaction: 
\begin{equation}\label{soi}
\hat{H}^{\text{SOI}} = \sum_{i,m} (-1)^m \: \hat{F}^{(1)}_{-m}(i) \: \hat{S}^{(1)}_{m}(i),
\end{equation}
which is a one-electron operator in the mean-field approximation and contains an orbital-dependent part $\hat{F}(i)$, a function of the position and momentum of electron $i$, and a spin-dependent part $\hat{S}(i)$, the spin operator of electron $i$; (ii) the Zeeman interaction:
\begin{equation}\label{zeeman}
\hat{H}^{\text{Z}} = \mu_B \sum_{i,m} (-1)^m \left( g_e B^{(1)}_{-m} \: \hat{S}^{(1)}_{m}(i) + B^{(1)}_{-m} \: \hat{L}^{(1)}_{m}(i) \right),
\end{equation}
which includes the interaction of the magnetic field $B$ with the electron spin, first term, and the interaction with the orbital motion of the electrons, second term. The sum over $i$ runs over all electrons, and the sum over $m$ runs over all spherical components of the spherical tensors of rank $(1)$. $g_e$ is the anomalous g-factor of the electron. Application of the Wigner-Eckart theorem to the matrix elements of the spin-orbit interaction gives:
\begin{equation}\label{soiwigner}
\begin{split}
H^{\text{SOI}}_{JI} &= \bra{jSM'} \hat{H}^{\text{SOI}} \ket{jSM} \\ &= \begin{aligned}[t] \sum_{m} (-1)^m & \bra{S1;Mm}\ket{S1;SM'} \\ &\frac{\bra{jS|} \sum_{i} \hat{F}^{(1)}_{-m}(i) \: \hat{S}^{(1)}(i) \ket{|iS}}{\sqrt{2S+1}} \end{aligned}\\ &= \sum_{m} (-1)^m S^{(1)}_{M'Mm} \frac{\bra{jSS} \sum_{i} \hat{F}^{(1)}_{-m}(i) \: \hat{S}^{(1)}_{0}(i) \ket{iSS}}{S} \\ &= i \sum_{\alpha} S_{M'M\alpha} \Im \frac{\bra{jSS} \sum_{i} \hat{F}_{\alpha}(i) \: \hat{S}_{z}(i) \ket{iSS}}{S} \\ &= i \sum_{\alpha} S_{M'M\alpha} h^{\text{SOI}}_{ji\alpha}.
\end{split}
\end{equation}
The second line of Eq.~(\ref{soiwigner}) results directly from the Wigner-Eckart theorem, where $\bra{S1;Mm}\ket{S1;SM'}$ is a Clebsch-Gordan coefficient in the nomenclature of Sakurai\cite{sakurai2014modern} and $\bra{jS|} \cdot \ket{|iS}$ denotes a reduced matrix element. In the third line, I express the Clebsch-Gordan coefficient using the matrix elements of the electron spin operator, and in the fourth line, I account for the purely imaginary nature of the resulting orbital-dependent operator. The last line of Eq.~(\ref{soiwigner}) provides the definition of $h^{\text{SOI}}_{ji\alpha}$. The sum over $m$ is a sum over the spherical components of the rank-1 spherical tensors as before, and the sum over $\alpha$ is the equivalent sum over the Cartesian components of the vector operators. Similar application of the Wigner-Eckart theorem to the Zeeman interaction reveals the following spin-component structure:
\begin{equation}\label{zeemanwigner}
\begin{split}
\hat{H}^{\text{Z}}_{JI} = &\bra{jSM'} \hat{H}^{\text{Z}} \ket{jSM} \\ = &\mu_B g_e \sum_{m} (-1)^m  B^{(1)}_{-m} \: S^{(1)}_{M'Mm} \delta_{ji} \\ &+ \mu_B \sum_{m} (-1)^m B^{(1)}_{-m} \: \bra{jSS} \hat{L}^{(1)}_{m} \ket{iSS} \delta_{M'M}. 
\end{split}
\end{equation}
Use of Eqs.~(\ref{soiwigner}) and (\ref{zeemanwigner}) in the second order term of Eq.~(\ref{expansion}), and keeping only the combination terms between the orbital Zeeman and spin-orbit interactions gives the second order contribution to the molecular g-tensor:
\begin{equation}\label{gtensorsoz}
\begin{split}
\Delta g^{(2)}_{\alpha\beta} &= \frac{1}{S^2} \sum_{j \ne 0} \frac{\Im \bra{0SS} \hat{L}_{\alpha} \ket{jSS} \Im \bra{jSS} \hat{F}_{\beta}\hat{S}_{z} \ket{0SS} }{E^{(0)}_j - E^{(0)}_i}  \\ & + \frac{1}{S^2} \sum_{j \ne 0} \frac{ \Im \bra{0SS} \hat{F}_{\alpha}\hat{S}_{z} \ket{jSS} \Im \bra{jSS} \hat{L}_{\beta} \ket{0SS}  }{E^{(0)}_j - E^{(0)}_i}.
\end{split}
\end{equation}
Equation~(\ref{gtensorsoz}) is the sum-over-states expression of the second order molecular g-tensor and demonstrates the relation between adiabatic terms in the molecular spin Hamiltonian and the traditional static spin Hamiltonian interactions. Similar account of the pure spin-orbit interaction terms in Eq.~(\ref{expansion}) gives the second-order sum-over-states expression for the molecular D-tensor, which also involves matrix elements of the spin-orbit interaction between spin manifolds that differ by one unit of angular momentum. 

Finally, the same-spin-manifold matrix elements of the first-order spin-orbit perturbing operator can be written using the Wigner-Eckart theorem as:
\begin{equation}\label{firstorderwfn}
\begin{split}
G^{(1)}_{JI} &= \bra{jSM'} \hat{G}^{(1)} \ket{iSM} = - \sum_{J \ne I} \frac{H^{\text{SOI}}_{JI}}{E^{(0)}_J - E^{(0)}_I} \\ &= -i S_{M'M\alpha} \sum_{j \ne i} \frac{h^{\text{SOI}}_{ji\alpha}}{E^{(0)}_j - E^{(0)}_i} = i S_{M'M\alpha} g^{(1)}_{ji\alpha},
\end{split}
\end{equation}
where in the last line I define the Cartesian components of the first-order perturbing operator $g_{ji\alpha}$. $g_{ji\alpha}$ are purely real functions and are symmetric matrices with respect to the orbital indeces.

\section{Proof of identity in Eq.~\ref{Dpot2}}\label{appendixb}

I prove in Appendix~\ref{appendixb} an identity that is needed to transform the second derivative non-adiabatic coupling matrix in Eq.~(\ref{Dpot2}):
\begin{equation}\label{Dpot3}
\begin{split}
    \bra{I'^{(0)}_{\textbf{R}}}& \partial_\mu \hat{G}^{(1)}_\textbf{R} \ket{\partial_\mu I^{(0)}_{\textbf{R}}} = \\ = &\sum_{J \ne I} \partial_\mu G_{I'J} \bra{J}\ket{\partial_\mu I} + \sum_{J \ne I} G_{I'J} \bra{\partial_\mu J}\ket{\partial_\mu I} + \\ &\sum_{K,J,J \ne K} \bra{I'}\ket{\partial_\mu K} G_{KJ} \bra{J}\ket{\partial_\mu I} \\ = &i S_{M'M\alpha} ( \sum_{j \ne i} \partial_\mu g_{ij\alpha} \bra{j}\ket{\partial_\mu i} + \sum_{j \ne i} g_{ij\alpha} \bra{\partial_\mu j}\ket{\partial_\mu i}  + \\ &\sum_{k,j,j \ne k} \bra{i}\ket{\partial_\mu k} g_{kj\alpha} \bra{j}\ket{\partial_\mu i} ) \\ = &i S_{M'M\alpha} ( \sum_{j \ne i} \bra{\partial_\mu i}\ket{j}  \partial_\mu g_{ji\alpha}  + \sum_{j \ne i} \bra{\partial_\mu i}\ket{\partial_\mu j} g_{ji\alpha} + \\ &\sum_{k,j, j \ne k} \bra{\partial_\mu i}\ket{j} g_{jk\alpha} \bra{\partial_\mu k}\ket{i}  ) \\ = &\bra{\partial_\mu I'^{(0)}_{\textbf{R}}} \partial_\mu \hat{G}^{(1)}_\textbf{R} \ket{I^{(0)}_{\textbf{R}}}
\end{split}
\end{equation}
The first line follows from the expansion of the first-order perturbing operator, the second line employs Eq.~(\ref{firstorderwfn}), and the third line uses the symmetry of $g_{ji\alpha}$ and of the derivative matrix elements in the case of real zero-order wavefunctions. 

\end{appendices}

\bibliography{dm.bib}

\end{document}